\documentclass[article,amsmath,amssymb,nofootinbib,aps,prc,superscriptaddress]{revtex4-2} 

\usepackage{graphicx}
\usepackage{epstopdf}
\usepackage{caption}
\usepackage{color}
\usepackage{bm}
\usepackage{textcomp} 
\usepackage{subfigure}
\usepackage{ulem}
\usepackage{color}
\usepackage{mathtools}
\usepackage{amsmath}

\usepackage{diagbox}
\usepackage{slashbox}
\usepackage{booktabs}
\usepackage{multirow}

\usepackage{array}
\usepackage{amssymb}

\graphicspath{{figures/}}

\begin{document}

\title{Investigation of Experimental Observables in Search of the Chiral Magnetic Effect in Heavy-ion Collisions in the STAR experiment}

\author{Subikash~Choudhury}
\affiliation{Fudan University, Shanghai, 200433}
\author{Xin~Dong}
\affiliation{Lawrence Berkeley National Laboratory, Berkeley, California 94720 }
\author{Jim~Drachenberg}
\affiliation{Abilene Christian University, Abilene, Texas   79699}
\author{James~Dunlop}
\affiliation{Brookhaven National Laboratory, Upton, New York 11973}
\author{ShinIchi~Esumi}
\affiliation{University of Tsukuba, Tsukuba, Ibaraki 305-8571, Japan}
\author{Yicheng~Feng}
\affiliation{Purdue University, West Lafayette, Indiana 47907}
\author{Evan~Finch}
\affiliation{Southern Connecticut State University, New Haven, Connecticut 06515}
\author{Yu~Hu}
\affiliation{Fudan University, Shanghai, 200433}
\affiliation{Brookhaven National Laboratory, Upton, New York 11973}
\author{Jiangyong~Jia}
\affiliation{Brookhaven National Laboratory, Upton, New York 11973}
\affiliation{State University of New York, Stony Brook, New York 11794}
\author{Jerome~Lauret}
\affiliation{Brookhaven National Laboratory, Upton, New York 11973}
\author{Wei~Li}
\affiliation{Rice University, Houston, Texas 77251} 
\author{Jinfeng~Liao}
\affiliation{
Physics Department and Center for Exploration of Energy and Matter, Indiana University, 2401 N Milo B. Sampson Lane, Bloomington, IN 47408, USA}
\author{Yufu~Lin}
\email{yufulin@mails.ccnu.edu.cn}
\affiliation{College of Physics and Technology, Guangxi Normal University, Guilin, 541004, China}
\affiliation{Central China Normal University, Wuhan, Hubei 430079, People’s Republic of China}
\author{Mike~Lisa}
\affiliation{Ohio State University, Columbus, Ohio 43210}
\author{Takafumi~Niida}
\affiliation{University of Tsukuba, Tsukuba, Ibaraki 305-8571, Japan}
\author{Robert~Lanny~Ray} 
\affiliation{University of Texas, Austin, Texas 78712}
\author{Masha~Sergeeva}
\affiliation{University of California, Los Angeles, California 90095}
\author{Diyu~Shen}
\email{dyshen@fudan.edu.cn}
\affiliation{Fudan University, Shanghai, 200433}
\author{Shuzhe~Shi}
\affiliation{Department of Physics, McGill University, 3600 University Street, Montreal, QC, H3A 2T8, Canada}
\author{Paul~Sorensen}
\affiliation{Brookhaven National Laboratory, Upton, New York 11973}
\author{Aihong~Tang}
\affiliation{Brookhaven National Laboratory, Upton, New York 11973}
\author{Prithwish~Tribedy}
\affiliation{Brookhaven National Laboratory, Upton, New York 11973}
\author{Gene~Van~Buren}
\affiliation{Brookhaven National Laboratory, Upton, New York 11973}
\author{Sergei~Voloshin}
\affiliation{Wayne State University, Detroit, Michigan 48201}
\author{Fuqiang~Wang}
\affiliation{Purdue University, West Lafayette, Indiana 47907}
\author{Gang~Wang}
\affiliation{University of California, Los Angeles, California 90095}
\author{Haojie~Xu}
\affiliation{Huzhou University, Huzhou, Zhejiang  313000}
\author{Zhiwan~Xu}
\affiliation{University of California, Los Angeles, California 90095}
\author{Nanxi~Yao}
\email{waclewacle@ucla.edu}
\affiliation{University of California, Los Angeles, California 90095}
\author{Jie~Zhao}
\affiliation{Purdue University, West Lafayette, Indiana 47907}



\begin{abstract}
The chiral magnetic effect (CME) is a novel transport phenomenon, arising from the interplay between quantum anomalies and strong magnetic fields  in chiral systems.
In high-energy nuclear collisions, the CME may survive the expansion of the quark-gluon plasma fireball and be detected in experiments. Over the past  two decades, the experimental searches for the CME have aroused extensive interest at the Relativistic Heavy Ion Collider (RHIC)  and the Large Hadron Collider (LHC). The main goal of this article is to investigate three pertinent experimental approaches: the $\gamma$ correlator, the $R$ correlator and the signed balance functions. We will exploit both simple Monte Carlo simulations and a realistic event generator (EBE-AVFD) to verify the equivalence in the  core components among these methods and to ascertain their sensitivities to the CME signal and the background contributions for the isobaric collisions at RHIC.
\begin{description}
\item[keywords]
chiral magnetic effect, anisotropic flow, heavy-ion collisions, quark-gluon plasma
\end{description}
\end{abstract}

\maketitle

\section{Introduction}

A system is called {\it chiral} if it is not invariant under mirror reflection.
The imbalance of right- and left-handed particles in a chiral system can
be quantified by chiral chemical potential ($\mu_5$). In a system of charged  fermions with  finite $\mu_5$, an electric current could 
be generated in the presence of a strong  magnetic field ($\overrightarrow{B}$),
\begin{equation}
\overrightarrow{J_e} \propto \mu_5\overrightarrow{B},
\label{eq:CME}
\end{equation}
which is theorized as the chiral magnetic
effect (CME)~\cite{Kharzeev_PLB2006,Kharzeev_NPA2008}. 
The CME physics encompasses a wide range of systems and has generated significant interdisciplinary interests (see some recent reviews in Refs.~\cite{Review1,Review2,Review3,Kharzeev:2020jxw}). 
For example, it  has been  observed in condensed matter systems
using Dirac and Weyl semimetals with emergent chiral quasiparticles (e.g., ZrTe$_{5}$~\cite{ZrTe5}, Na$_3$Bi~\cite{Na3Bi}, TaAs~\cite{TaAs} and TaP~\cite{TaP}). In this article, we present method studies in
search of the CME in high-energy nuclear collisions.

Ultra-relativistic heavy-ion  collisions have been performed in the experimental facilities, such as the Relativistic Heavy Ion Collider (RHIC) and
the Large Hadron Collider (LHC). These experiments aim to create a new phase of hot and 
dense nuclear matter with a temperature above several trillion
Kelvin, consisting of deconfined quarks and gluons, called the quark-gluon plasma (QGP)~\cite{STAR_white,PHENIX_white,PHOBOS_white,BRAHMS_white}. Inside the QGP, the chiral symmetry is approximately restored, and light-flavor quarks $u$ and $d$ become nearly massless and hence chiral.
The two preconditions (finite $\mu_5$ and $\overrightarrow{B}$ field) for the CME could be realized in heavy-ion collisions as follows. 
\begin{itemize}
\item According to quantum chromodynamics (QCD), if the topological solutions of the  SU(3) gauge group are chiral, they can transfer chirality to quarks via the chiral anomaly~\cite{ChiralAnomality1,ChiralAnomality2},
forming local chiral domains with finite $\mu_5$ in a QGP~\cite{Kharzeev_PLB2006,Kharzeev_NPA2008,Kharzeev_NPA2007,Kharzeev_PLB2002,Yin_PRL2015,Kharzeev_PRL2010}. Note that the global chirality imbalance still vanishes when averaged over   an infinite number of domains.
\item In non-central heavy-ion collisions, extremely strong magnetic fields ($B \sim 10^{14}$~T) can be formed~\cite{Kharzeev_PLB2006,Kharzeev_NPA2007}, by energetic protons. (By convention, the participating nucleons in the overlap region are called participants, and the rest, spectators.)
\end{itemize}
Therefore, the  CME could happen in heavy-ion collisions, and cause an electric current along the $\overrightarrow{B}$ direction. Since $\overrightarrow{B}$ is approximately perpendicular to the reaction plane ($\Psi_{\rm RP}$) that contains the impact
parameter and the beam momenta of a collision,  the CME will manifest as an electric charge transport phenomenon across the reaction plane.

In view of the CME-induced charge transport and other modes of collective motion of the QGP, the azimuthal distribution of particles is often Fourier-decomposed for given transverse momentum ($p_T$) and pseudorapidity ($\eta$) in an event:
\begin{equation}
    \frac{dN_{\alpha}}{d\phi^*} \approx \frac{N_\alpha}{2\pi} [1 + 2v_{1,\alpha}\cos(\phi^*) + 2v_{2,\alpha}\cos(2\phi^*) + 2v_{3,\alpha}\cos(3\phi^*) + ... + 2a_{1,\alpha}\sin(\phi^*) + ...],
\label{equ:Fourier_expansion}
\end{equation}
\noindent where $\phi$ is the azimuthal angle of a particle, and $\phi^* = \phi - {\rm \Psi_{RP}}$.
The subscript $\alpha$ ($+$ or $-$) denotes the charge sign of a particle.
Conventionally, the coefficients $v_1$, $v_2$ and $v_3$ are called ``directed flow", ``elliptic flow", and ``triangular flow", respectively. They 
reflect the hydrodynamic response of the QGP medium to the initial collision geometry and to its fluctuations~\cite{HYDRO_review}.
Here ``RP" does not necessarily mean the reaction plane, but rather a flow symmetry plane obtained from the collective motion of final particles. For simplicity, we still use RP in the following discussions, and RP denotes a specific flow plane.
The coefficient $a_1$ (with $a_{1,-} = -a_{1,+}$) characterizes the electric charge separation with respect to the flow plane, e.g., due to the CME.
It is tempting to average the $a_{1,\pm}$
coefficient in Eq.~\ref{equ:Fourier_expansion} over events to measure the CME signal. 
However, since $\mu_5$ flips sign on a domain-by-domain basis with equal probability (the global chirality  should be balanced), the event-averaged $a_{1,\pm}$ is zero by construction. 
Therefore, most experimental observables have been designed to measure the $a_{1,\pm}$  fluctuations  
with respect to the flow plane.

The experimental confirmation of the CME in heavy-ion collisions
will simultaneously advance several frontiers of fundamental physics: 
the evolution of the strong magnetic field, the topological
phases of QCD, and the chiral symmetry restoration of
strong interactions. A community-wide interest has been drawn to search for the CME at RHIC~\cite{STAR1,STAR2,STAR3,STAR4,STAR5,STAR6,UU1,Prithwish,JieZhao,PHENIX1,PHENIX2} and
the LHC~\cite{ALICE,CMS1,CMS2,ALICE2, ALICE3} over the past two decades. However, there is no definite conclusion so far for the existence of the CME in heavy-ion collisions. The main challenge in the CME search lies in the background contributions to the experimental observables. For example, mechanisms such as resonances with finite elliptic flow can also enhance charge fluctuations across  flow planes~\cite{Asakawa:2010bu,Liao:2010nv,Bzdak:2010fd,Petersen:2010di,Bzdak:2012ia,Pratt:2010zn,Schlichting:2010qia,Wang:2009kd,Wang:2016iov,Feng:2018chm}. To better gauge the background, the STAR experiment at RHIC has collected a large data sample of isobaric collisions, and performed the corresponding CME-related analyses.

The two isobaric systems, namely $^{96}_{44}$Ru + $^{96}_{44}$Ru and $^{96}_{40}$Zr + $^{96}_{40}$Zr, have the same number of nucleons and hence similar amounts of elliptic flow, but different numbers of protons, which causes a difference in the magnetic field strength and in turn, a difference in the CME signal~\cite{UU_theory,isobar1,isobar2}.
By keeping the background unchanged and varying the signal level, the two isobaric systems provide an ideal test ground for the CME study.
The STAR Collaboration has implemented a blind-analysis recipe~\cite{STARBlinding} to eliminate unintentional biases in data analyses, and currently all the analysis codes have been frozen as part of the blinding procedure.
Since the isobaric collisions will be examined with multiple observables, it is desirable to learn the connection and the difference between them, as well as their  sensitivities to the CME signals. The objective of this paper is to do apples-to-apples comparisons between method kernels. For the sensitivity test of the final observables being used in the STAR  blind analyses, we shall apply the STAR frozen codes to simulated events with various CME inputs. This work will provide an important reference point for the interpretation of the  isobaric-collision data coming out of the blind analyses.

The paper is structured as follows. In Sec.~II, we  focus on three observables in search of the CME: the $\gamma$ correlator~\cite{Sergei2004}, the $R$ correlator~\cite{PHENIX2,RCorr-2018} and the signed balance functions~\cite{SBF-Aihong-2020,Yufu-2020}.
We shall uncover the relations between them. Section~III describes the simple Monte Carlo calculations, as well as a more realistic event generator: 
Event-By-Event Anomalous-Viscous Fluid Dynamics (EBE-AVFD) ~\cite{Shi:2017cpu,Jiang:2016wve,Shi:2019wzi}.
These simulations are deployed in Sec.~IV to compare the core components of the experimental observables. In Sec.~V, EBE-AVFD is used to estimate the sensitivities of the final observables to the CME signal for the isobaric collisions under study. Section VI gives the summary and outlook.

\section{Experimental Observables}
\label{Sec.II}
Several observables have been proposed to search for the CME in heavy-ion collisions, including the $\gamma$ correlator~\cite{Sergei2004}, 
the $R(\Delta S_m)$ correlator ~\cite{PHENIX2,RCorr-2018}, and the signed balance functions~\cite{SBF-Aihong-2020,Yufu-2020}. It is not a surprise that these methods provide
largely overlapping information, since they all 
make use of similar inputs of particle azimuthal correlations. We will review these approaches, and reveal the relations among them.

\subsection{$\gamma$ Correlator}

The three-point $\gamma$ correlator (later specified as $\gamma_{112}$) measures the fluctuations of
charge separations or $a_{1,\pm}$ coefficients
with respect to a flow plane~\cite{Sergei2004},
\begin{eqnarray}
\gamma_{112} &\equiv&  \langle \cos(\phi_\alpha + \phi_\beta -2{\rm \Psi_{RP}}) \rangle \nonumber \\
&=& \langle\cos(\phi^*_{\alpha})\cos(\phi^*_{\beta}) -
\sin(\phi^*_{\alpha})\sin(\phi^*_{\beta})\rangle \nonumber \\
&=& (\langle v_{1,\alpha}v_{1,\beta}\rangle + B_{\rm IN}) -(\langle a_{1,\alpha}a_{1,\beta}\rangle + B_{\rm OUT}), \label{eq:ThreePoint}
\end{eqnarray}
\noindent where the averaging is done over all combinations of particles $\alpha$ and $\beta$ in an event and over all events.
The trigonometric expansion exhibits the difference between {\it in-plane} and {\it out-of-plane} projections of azimuthal correlations.
The third term in Eq.~\ref{eq:ThreePoint}, $\langle a_{1,\alpha}a_{1,\beta}\rangle$,
represents the  fluctuations of $a_{1,\pm}$ coefficients,  the main target of the CME search. Other terms are presumably irrelevant to the CME: $\langle v_{1,\alpha}v_{1,\beta}\rangle$ is related to directed flow, and is expected to be charge-independent and unrelated to 
the electromagnetic field in symmetric A+A collisions; $B_{\rm IN}$ and $B_{\rm OUT}$  represent other
possible background correlations in and out of the flow plane, respectively.
When we take the difference between opposite-sign and same-sign $\gamma_{112}$ correlators,
\begin{equation}
\Delta \gamma_{112} \equiv \gamma^{\rm OS}_{112} - \gamma^{\rm SS}_{112}, 
\end{equation}

\noindent the $\langle v_{1,\alpha}v_{1,\beta}\rangle$ terms cancel out, as well as a large portion of ($B_{\rm IN}-B_{\rm OUT}$).
A residual flow-plane dependent background in ($B_{\rm IN}-B_{\rm OUT}$) still exists at a level proportional to elliptic flow, which is the major background source in the $\Delta \gamma_{112}$ measurements. 
In practice, the flow plane is approximated with the ``event plane" ($\rm \Psi_{EP}$) reconstructed with detected particles,
and then the measurement is corrected for the finite event plane resolution~\cite{Methods}.
The main advantages of $\gamma_{112}$ lie in its direct connection to $a_1$ and a straightforward procedure that corrects for the finite event plane resolution.

The flow-plane-related backgrounds in $\Delta\gamma_{112}$ can be understood with the example of resonance decays.
As resonances flow with the QGP medium, their decay daughters generate random (if parents possess no global spin alignment),
event-by-event charge separation across the flow plane~\cite{Wang:2009kd,Schlichting:2010qia}.
The flowing resonance picture can be generalized to  transverse momentum conservation (TMC)~\cite{Pratt:2010zn,Bzdak:2012ia} 
and local charge conservation (LCC)~\cite{Schlichting:2010qia}.
Ideally, the two-particle correlator, 
\begin{eqnarray}
\delta &\equiv& \langle \cos(\phi_\alpha -\phi_\beta) \rangle \nonumber \\
&=& (\langle v_{1,\alpha}v_{1,\beta}\rangle + B_{\rm IN}) +(\langle a_{1,\alpha}a_{1,\beta}\rangle + B_{\rm OUT}),
\label{eq:delta}
\end{eqnarray}
should also reflect $\langle a_{1,\alpha} a_{1,\beta} \rangle$,
but in reality it is dominated by short-range two-particle correlations.
For example, the TMC effect contributes the following pertinent correlations in $\Delta \delta$ and $\Delta \gamma_{112}$~\cite{Bzdak:2012ia}:
\begin{eqnarray}
\Delta \delta^{\rm TMC} &\approx& -\frac{1}{N}
\frac{\langle p_T \rangle^2_{\rm \Omega}}{\langle p_T^2 \rangle_{\rm F}}
\frac{1+({\bar v}_{2,{\rm \Omega}})^2-2{\bar{\bar v}}_{2,{\rm F}}{\bar v}_{2,{\rm \Omega}}} {1-({\bar{\bar v}}_{2,{\rm F}})^2},
\label{eq:TMC1}
\\
\Delta \gamma^{\rm TMC}_{112} &\approx& -\frac{1}{N}
\frac{\langle p_T \rangle^2_{\rm \Omega}}{\langle p_T^2 \rangle_{\rm F}}
\frac{2{\bar v}_{2,{\rm \Omega}}-{\bar{\bar v}}_{2,{\rm F}}-{\bar{\bar v}}_{2,{\rm F}}({\bar v}_{2,{\rm \Omega}})^2} {1-({\bar{\bar v}}_{2,{\rm F}})^2}
\nonumber \\
&\approx& \kappa^{\rm TMC}_{112} \cdot v_{2,{\rm \Omega}} \cdot \Delta \delta^{\rm TMC},
\label{eq:TMC2}
\end{eqnarray}
where $\kappa^{\rm TMC}_{112} = (2{\bar v}_{2,{\rm \Omega}}-{\bar{\bar v}}_{2,{\rm F}})/v_{2,{\rm \Omega}}$,
and ${\bar v}_{2}$ and ${\bar{\bar v}}_{2}$ represent the $p_T$- and $p_T^2$-weighted moments of $v_2$, respectively.
The subscript ``F" denotes an average over all produced particles in the full phase space, whereas
the actual measurements only sample a fraction of the full space, denoted by ``${\rm \Omega}$".
The background contribution due to the LCC effect has a similar characteristic structure
as in Eqs.~\ref{eq:TMC1} and \ref{eq:TMC2}~\cite{Pratt:2010zn,Schlichting:2010qia}. This motivates a normalization of $\Delta \gamma$ by $v_2$ and $\Delta \delta$:
\begin{equation}
    \kappa_{112} \equiv \frac{\Delta \gamma_{112}}{v_2 \cdot \Delta \delta}.
\label{kappa112}
\end{equation}
A CME signal will make $\kappa_{112}$  larger than $\kappa^{\rm TMC/LCC}_{112}$. While a reliable estimate of $\kappa^{\rm TMC/LCC}_{112}$ is still elusive, the comparison of $\Delta\gamma_{112}$ (and $\kappa_{112}$) between isobaric collisions may give a more definite conclusion on the CME signal.

It is intuitive to introduce some derivative $\gamma$ correlators and the corresponding $\kappa$ observables~\cite{CMS2}, for example,
\begin{eqnarray}
\gamma_{123} \equiv \langle \langle \cos(\phi_\alpha + 2\phi_\beta -3{\rm \Psi_{3}}) \rangle\rangle&,~\ & \kappa_{123} \equiv \frac{\Delta \gamma_{123}}{v_3 \cdot \Delta \delta}, 
\\
\gamma_{132} \equiv \langle \langle \cos(\phi_\alpha - 3\phi_\beta + 2{\rm \Psi_{2}}) \rangle\rangle&,~\ &
\kappa_{132} \equiv \frac{\Delta \gamma_{132}}{v_2 \cdot \Delta \delta},
\end{eqnarray}
where $\Psi_{2}$ and $\Psi_{3}$ represent the $2^{\rm nd}$- and $3^{\rm rd}$-order flow planes, respectively.
However, these observables may not serve as good  background estimates for $\gamma_{112}$ or $\kappa_{112}$. 
Background-only AMPT calculations indeed show that $\kappa_{123}$ and $\kappa_{132}$ are not equal to $\kappa_{112}$ in Au+Au collisions at 
200 GeV~\cite{Subikash}.
Therefore, in the following sections we will not extend our method study to these derivative correlators.

\subsection{$R$ correlator}
The $R(\Delta S_m)$ correlator~\cite{PHENIX2,RCorr-2018} takes the double ratio of four event-by-event distributions,
\begin{equation}
R(\Delta S_m) \equiv \frac{N(\Delta S_{m,\rm real})}{N(\Delta S_{m,\rm shuffled})} / \frac{N(\Delta S^{\perp}_{m,\rm real})}{N(\Delta S^{\perp}_{m,\rm shuffled})}, ~\ m = 2,3,...,
\end{equation}
where in a real event the charge separation perpendicular to the $m^{\rm th}$-order flow plane ($\Psi_m$) is expressed as 
\begin{equation}
\Delta S_{m,\rm real} = \langle\sin(\frac{m}{2}\phi^*_m)\rangle_{N_+} - \langle\sin(\frac{m}{2}\phi^*_m)\rangle_{N_-}.   
\end{equation}
Here $\phi^*_m = \phi - {\rm \Psi}_m$, and $N_+$($N_-$) is the number of positively (negatively) charged particles in a specific event. Weighted averages could be used to take into account the azimuthal acceptance of the detector.  $\Delta S^{\perp}_m$ denotes the  charge separation parallel to $\Psi_m$, and is defined similar to $\Delta S_m$, except that ${\rm \Psi}_m$ is replaced with $({\rm \Psi}_m + \pi/m)$ to provide a baseline unrelated to the magnetic field.
The $\Delta S^{(\perp)}_{m,\rm shuffled}$ distributions are obtained via random charge reassignment (shuffling) of the reconstructed tracks in each real event, while respecting the multiplicities of positive and negative charges. Ideally the CME will cause a concave shape in the double ratio of the $R(\Delta S_2)$ distribution, which presumably differs from the $R(\Delta S_3)$ shape~\cite{RCorr-2018}. The latter is intended as a background estimate similar to the role of $\gamma_{123}$.

The $R(\Delta S_2)$ distribution is fit with a Gaussian(inverse-Gaussian) function, if it bears a convex(concave) shape, and the Gaussian width ($\sigma_{R2}$) is used to reflect the CME signal. Since the four core components, $\Delta S^{(\perp)}_{2,\rm real(shuffled)}$,  all roughly follow Gaussian distributions, we can establish the relation between $\sigma_{R2}$ and the RMS values of the core-component distributions:
\begin{equation}
\frac{S_{\rm concavity}}{\sigma_{R2}^2}   = \frac{1}{\langle (\Delta S_{2,\rm real})^2 \rangle} - \frac{1}{\langle (\Delta S_{2,\rm shuffled})^2\rangle} - \frac{1}{\langle (\Delta S_{2,\rm real}^{\perp})^2  \rangle} + \frac{1}{\langle (\Delta S_{2,\rm shuffled}^{\perp})^2  \rangle}.
\label{eq:sigma_R2} 
\end{equation}
The sign of concavity, $S_{\rm concavity}$, is positive if the double ratio is convex, and negative  for the concave shape. We shall first understand the meaning of each term on the right. For simplicity, we use unity weights, and the first term can be expanded as
\begin{eqnarray}
& &(\Delta S_{2,\rm real})^2 \nonumber \\
&\equiv& (\frac{\sum_{i=1}^{N_+}\sin(\phi^*_i)}{N_+}-\frac{\sum_{i=1}^{N_-}\sin(\phi^*_i)}{N_-})^2 \nonumber \\
&=& \frac{\sum_{i=1}^{N_+}\sin^2(\phi^*_i)+\sum_{i\neq j}^{N_+}\sin(\phi^*_i)\sin(\phi^*_j)}{N_+^2}+\frac{\sum_{i=1}^{N_-}\sin^2(\phi^*_i)+\sum_{i\neq j}^{N_-}\sin(\phi^*_i)\sin(\phi^*_j)}{N_-^2}-\frac{2\sum_{i=1,j=1}^{N_+,N_-}\sin(\phi^*_i)\sin(\phi^*_j)}{N_+N_-} \nonumber \\
&=& \frac{\langle \sin^2(\phi^*_i)\rangle_{N_+} +(N_+-1) \langle \sin(\phi^*_i)\sin(\phi^*_j)\rangle_{N_+N_+}}{N_+}+\frac{\langle \sin^2(\phi^*_i)\rangle_{N_-} +(N_--1) \langle \sin(\phi^*_i)\sin(\phi^*_j)\rangle_{N_-N_-}}{N_-} \nonumber \\
& &-2\langle \sin(\phi^*_i)\sin(\phi^*_j) \rangle_{N_+N_-}
\label{eq:dS}
\end{eqnarray}
Using the trigonometric identities, $\sin^2(x) = [1-\cos(2x)]/2$ and $2\sin(x)\sin(y) = \cos(x-y)-\cos(x+y)$, we have the average of Eq.~\ref{eq:dS} over all events:
\begin{eqnarray}
\langle(\Delta S_{2,\rm real})^2\rangle 
&=& \frac{1-v_2^+}{2N_+} + \frac{N_+-1}{2N_+}(\delta^{++}-\gamma^{++}_{112})+\frac{1-v_2^-}{2N_-} + \frac{N_--1}{2N_-}(\delta^{--}-\gamma^{--}_{112}) - (\delta^{+-}-\gamma^{+-}_{112}) \\
&\approx& \frac{2(1-v_2-\delta^{\rm SS}+\gamma_{112}^{\rm SS})}{M} - \Delta\delta + \Delta\gamma_{112}.
\label{eq:s-expand}
\end{eqnarray}
The last line assumes $v_2^+ \approx v_2^-$ and $N_+ \approx N_- = M/2$. Even before the approximations are taken, it is clear that $\langle(\Delta S_{2,\rm real})^2\rangle$ can be expressed by $N_{+(-)}$, $v_2$, $\delta$ and $\gamma_{112}$. Similarly,
\begin{eqnarray}
\langle(\Delta S_{2,\rm real}^\perp)^2\rangle 
&=& \frac{1+v_2^+}{2N_+} + \frac{N_+-1}{2N_+}(\delta^{++}+\gamma^{++}_{112})+\frac{1+v_2^-}{2N_-} + \frac{N_--1}{2N_-}(\delta^{--}+\gamma^{--}_{112}) - (\delta^{+-}+\gamma^{+-}_{112}) \\
&\approx& \frac{2(1+v_2-\delta^{\rm SS}-\gamma_{112}^{\rm SS})}{M} - \Delta\delta - \Delta\gamma_{112}.
\label{eq:sp-expand}
\end{eqnarray}
For the shuffled terms, $v_2^+$ and $v_2^-$ will be roughly replaced with $(v_2^+ + v_2^-)/2$, all the $\delta$ correlators will be replaced with $(\delta^{\rm OS}+\delta^{\rm SS})/2$, and all the $\gamma$ correlators, with $(\gamma^{\rm OS}+\gamma^{\rm SS})/2$.
Therefore, we have
\begin{eqnarray}
\langle(\Delta S_{2,\rm shuffled})^2\rangle 
&\approx& \frac{2(1-v_2)-\delta^{\rm SS}-\delta^{\rm OS}+\gamma_{112}^{\rm SS}+\gamma_{112}^{\rm OS}}{M}
\label{eq:s-S-expand}\\
\langle(\Delta S_{2,\rm shuffled}^\perp)^2\rangle 
&\approx& \frac{2(1+v_2)-\delta^{\rm SS}-\delta^{\rm OS}-\gamma_{112}^{\rm SS}-\gamma_{112}^{\rm OS}}{M}.
\label{eq:sp-S-expand}
\end{eqnarray}

It is appealing to construct a double-subtraction observable based on the four core components of $R(\Delta S_2)$, which shows an apparent link to $\Delta\gamma_{112}$:
\begin{equation}
\Delta_{R2} \equiv \langle (\Delta S_{2,\rm real})^2 \rangle - \langle (\Delta S_{2,\rm shuffled})^2 \rangle - \langle (\Delta S^{\perp}_{2,\rm real})^2 \rangle  + \langle (\Delta S^{\perp}_{2,\rm shuffled})^2 \rangle \approx 2(1-\frac{1}{M})\Delta \gamma_{112}.
\label{eq:relation1}
\end{equation}
This relation will be tested in Sec.~IV. To make the connection between the final observable, $\sigma^2_{R2}$ and $\Delta\gamma$, we put Eqs.~\ref{eq:s-expand}, ~\ref{eq:sp-expand},~\ref{eq:s-S-expand} and~\ref{eq:sp-S-expand} into Eq.~\ref{eq:sigma_R2}, and reach
\begin{equation}
\frac{S_{\rm concavity}}{\sigma_{R2}^2} \approx -\frac{M}{2}(M-1)\Delta\gamma_{112}.    
\label{eq:relation2}
\end{equation}
Here we further assume that in each of the four $\langle (\Delta S_2)^2\rangle$ terms, $\frac{2}{M}$ is much larger than other contributions, among which $\Delta\delta$ typically has the largest magnitude. This assumption may not always hold true, depending on the collision details such as beam energy and centrality.
The relation in Eq.~\ref{eq:relation2} implies that if $\Delta\gamma$ is positively (negatively) finite, the $R(\Delta S_2)$ distribution will exhibit a concave(convex) shape, and vice versa. 

The above derivation is with respect to the known flow plane $\Psi_m$,  which is not precisely known experimentally, and is assessed by the reconstructed event plane with a finite resolution. 
The empirical correction for the event plane resolution used in experiment for $\Delta_{R2}$ and $\sigma^2_{R2}$~\cite{Magdy:2017yje} is nontrivial. An analytical resolution correction has been derived in Ref.~\cite{Feng:2020cgf}. 
Eqs.~\ref{eq:relation1} and \ref{eq:relation2} can provide approximate approaches to examine the resolution correction factor. 

Multiplicity fluctuations widen the $R(\Delta S)$ distributions. To reduce this effect to $R(\Delta S_2)$, a scaling was proposed in Ref.~\cite{RCorr-2018}, used in experimental data analysis, and included in the frozen code of the STAR isobar blind analysis.
The $R(\Delta S_2)$ distribution is converted into the
$R(\Delta S^{'}_2)$ distribution by dividing the horizontal axis by the RMS of the $N(\Delta S_{2, \rm shuffled})$ distribution, i.e.,  $\Delta S_2^{'} = \Delta S/\sqrt{\langle(\Delta S_{2,\rm shuffled})^2\rangle}$ (see Sec.\ref{sectV} for details).
Then, the width of $R(\Delta S^{'}_2)$ becomes 
\begin{equation}
\frac{S_{\rm concavity}}{\sigma_{R2'}^2} = 
\frac{S_{\rm concavity}}{\sigma_{R2}^2} \langle(\Delta S_{2,\rm shuffled})^2\rangle 
\approx -\frac{M}{2}(M-1)\Delta\gamma_{112} \times \frac{2}{M} 
\approx -M \Delta\gamma_{112} .
\end{equation}

\subsection{Signed Balance Functions}

Another method was recently proposed and invokes signed balance functions~\cite{SBF-Aihong-2020},
\begin{eqnarray}
\Delta B_y 
&\equiv& \Big[\frac{N_{y(+-)}-N_{y(++)}}{N_+} - \frac{N_{y(-+)}-N_{y(--)}}{N_-}\Big] - \Big[\frac{N_{y(-+)}-N_{y(++)}}{N_+} - \frac{N_{y(+-)}-N_{y(--)}}{N_-}\Big] \nonumber \\
&=& \frac{N_+ + N_-}{N_+N_-}[N_{y(+-)} - N_{y(-+)}].
\label{eq:by}
\end{eqnarray}
where $N_{y(\alpha\beta)}$ is an event-by-event quantity, and denotes the number of pairs within which particle $\alpha$ is ahead of particle $\beta$  in the direction perpendicular to the reaction plane ($p_y^\alpha > p_y^\beta$). Similarly, we can construct a $\Delta B_x$ to count the number of pairs along the in-plane direction. 
Then the final observable is based on the widths of the $\Delta B_y$ and the $\Delta B_x$ distributions:
\begin{equation}
r \equiv \sigma(\Delta B_y) / \sigma(\Delta B_x).
\label{rlab}
\end{equation}
Intuitively, the CME will lead to $r>1$, since the CME-induced charge separation will cause more fluctuations of pair ordering  across the reaction plane.
The ratio $r$ can be calculated in both the laboratory frame
($r_{\rm lab}$) and the pair's rest frame ($r_{\rm rest}$). It is argued that the
rest frame is the most appropriate frame for $r$ to study charge separations, and the further ratio,
\begin{equation}
R_{\rm B} = r_{\rm rest} / r_{\rm lab},
\label{eq:RB}
\end{equation}
can help differentiate the background from the real CME signal~\cite{SBF-Aihong-2020}. 
Extra care is also needed to correct the $r$ observables for the event plane resolution.
 
In each event, we can rewrite the core component of $\Delta B_y$ as follows,
\begin{equation}
N_{y(\alpha\beta)}-N_{y(\beta\alpha)} = \sum_{\alpha,\beta} {\rm Sign}[p_{T,\alpha}\sin(\phi^*_\alpha)-p_{T,\beta}\sin(\phi^*_\beta)].    
\end{equation}
Compared with other methods, the signed balance functions could be more sensitive to the local CME domains that move with the expanding medium, because this method takes into account the transverse-momentum ordering instead of only the azimuthal angle. For example, a pair of particles going in the same direction can still be regarded as a case of charge separation by the signed balance functions, but not by other methods that only consider the azimuthal angle. This advantage, however, probably will not make a prominent difference, if the local domains merge into a global charge separation for the whole event after a full hydrodynamic evolution. Thus, to make connection to other observables, we take the first approximation by replacing $p_T$ with mean $p_T$, so that $p_T$ can be dropped out for the time being, and  only the azimuthal angle is exploited as done in other methods. Next, we want to unpack the Sign() function, and directly use $[\sin(\phi^*_\alpha) - \sin(\phi^*_\beta)]$, which  requires a normalization factor, $C_y$. In view of the event average, we have
\begin{eqnarray}
\langle N_{y(\alpha\beta)}-N_{y(\beta\alpha)} \rangle &\approx& C_y \Big\langle \sum_{\alpha,\beta} [\sin(\phi^*_\alpha) - \sin(\phi^*_\beta)] \Big\rangle \nonumber \\
&=& C_y \Big\langle [N_\beta \sum_{\alpha}\sin(\phi^*_\alpha) - N_\alpha \sum_{\beta}\sin(\phi^*_\beta)] \Big\rangle \nonumber \\
&=& C_y N_\alpha N_\beta \langle \langle \sin( \phi^*)\rangle _{N_\alpha}-\langle \sin(\phi^*)\rangle_{N_\beta} \rangle. \label{eq:constant}
\end{eqnarray}
The constant can be determined by explicitly counting the pairs, with $\frac{dN}{d\phi^*}$ from Eq.~\ref{equ:Fourier_expansion}.
\begin{eqnarray}
\langle N_{y(\alpha\beta)}-N_{y(\beta\alpha)} \rangle &=& 2\int_{-\pi/2}^{\pi/2} 
\Big[\int_{-\pi/2}^{\phi^*_\alpha} \frac{dN}{d\phi^*_\beta}d\phi^*_\beta+\int_{\pi-\phi^*_\alpha}^{3\pi/2} \frac{dN}{d\phi^*_\beta}d\phi^*_\beta-\int^{\pi-\phi^*_\alpha}_{\phi^*_\alpha} \frac{dN}{d\phi^*_\beta}d\phi^*_\beta\Big]\frac{dN}{d\phi^*_\alpha}d\phi^*_\alpha \nonumber \\
&\approx& \frac{8}{\pi^2}(1+\frac{2}{3}v_2)N_\alpha N_\beta (a_{1,\alpha}-a_{1,\beta}).  \label{eq:integral_result}  
\end{eqnarray}
By comparing Eqs.~\ref{eq:constant} and \ref{eq:integral_result}, we learn $C_y = 8(1+2v_2/3)/\pi^2$. Therefore,
if we ignore the $p_T$ weight,  $\langle\Delta B_y\rangle$ becomes $\frac{8M}{\pi^2}(1+\frac{2}{3}v_2) \langle\langle \sin(\phi^*) \rangle_{N_+} -\langle \sin(\phi^*)\rangle_{N_-} \rangle$, which displays a function form akin to $\langle S_{2,\rm real}\rangle$. 
In a similar way,
we assume
\begin{equation}
\langle N_{x(\alpha\beta)}-N_{x(\beta\alpha)} \rangle \approx  C_x N_\alpha N_\beta \langle\langle \cos(\phi^*) \rangle _{N_\alpha}-\langle \cos(\phi^*)\rangle_{N_\beta} \rangle,  
\end{equation}
and the explicit counting gives
\begin{eqnarray}
\langle N_{x(\alpha\beta)}-N_{x(\beta\alpha)} \rangle &=& 2\int_{0}^{\pi} 
\Big[\int_{\phi^*_\alpha}^{2\pi-\phi^*_\alpha} \frac{dN}{d\phi^*_\beta}d\phi^*_\beta-\int^{\phi^*_\alpha}_{-\phi^*_\alpha} \frac{dN}{d\phi^*_\beta}d\phi^*_\beta\Big]\frac{dN}{d\phi^*_\alpha}d\phi^*_\alpha \nonumber \\
&\approx& \frac{8}{\pi^2}(1-\frac{2}{3}v_2)N_\alpha N_\beta (v_{1,\alpha}-v_{1,\beta}).   
\end{eqnarray}
Thus, $C_x = 8(1-2v_2/3)/\pi^2$, and 
$\langle\Delta B_x\rangle$ becomes  $\frac{8M}{\pi^2}(1-\frac{2}{3}v_2) \langle\langle \cos(\phi^*)\rangle_{N_+} -\langle \cos(\phi^*)\rangle_{N_-} \rangle$, resembling $\langle \Delta S_{2,\rm real}^{\perp}\rangle$. 

In reality, both $\langle\Delta B_{y(x)}\rangle$
and $\langle \Delta S_{2,\rm real}^{(\perp)}\rangle$ are zero,
but the derivation of $C_y$ and $C_x$ provides  insight into the meaning of the signed balance functions.
Our goal is to relate the RMS values of the $\Delta B_y$ and $\Delta B_x$ distributions to 
the other observables.
Here we directly give the following relations, with the details of the analytical derivations explained in Appendix~\ref{appendix1}.
\begin{eqnarray}
\sigma^2(\Delta B_y) &\approx& \frac{4M}{3}+ \frac{64M^2}{\pi^4}(1+ \frac{4}{3} v_2)(a_{1,+}-a_{1,-})^2 
\\
\sigma^2(\Delta B_x) &\approx& \frac{4M}{3}+ \frac{64M^2}{\pi^4}(1-\frac{4}{3}v_2)(v_{1,+}-v_{1,-})^2 .
\end{eqnarray}
Then we define an observable that further connects the signed balance function to the $\gamma$ correlator:
\begin{equation}
\Delta_{\rm SBF} \equiv \sigma^2(\Delta B_y) - \sigma^2(\Delta B_x) \approx  \frac{128M^2}{\pi^4}(\Delta\gamma_{112}-\frac{4}{3}v_2\Delta\delta).   \label{eq:relation3} 
\end{equation}
Note that the signed balance functions take into account not only the azimuthal angles of particles, but also their momenta. 
If the ratio definition in Eq.~\ref{rlab} is transformed to $\sigma^2(\Delta B_y) - \sigma^2({\Delta B_x})$, then this method is roughly equivalent to $(\Delta \gamma_{112}-\frac{4}{3}v_2\Delta\delta)$ with momentum weighting.

\section{Model descriptions}
We now describe a toy model that undertakes simple Monte Carlo calculations, as well as the more realistic event generator,  EBE-AVFD~\cite{Shi:2017cpu,Jiang:2016wve,Shi:2019wzi}.

\subsection{Toy Model}
We invoke a toy model to verify the mathematical relation between different methods with simplified Monte Carlo calculations.
In this setup, particle spectra, collective flow and charge-separation signals are well constrained and  conveniently adjusted, such that the response of each method to the signal and the background can be understood in a controlled manner.  In the simulations, each event consists of 195 $\pi^+$ and 195 $\pi^-$ mesons to match the total multiplicity at midrapidities within 2 units of rapidity in 30-40\% central Au+Au collisions at $\sqrt{s_{NN}}=$ 200 GeV~\cite{STAR-pion-spct}. In the background-free case, all pions are treated as primordial ones (none from resonance decays), and their azimuthal angle distribution is configured according to Eq. \ref {equ:Fourier_expansion} with $a_1$, $v_2$ and $v_3$. The elliptic flow ($v_2$) is introduced by an NCQ-inspired function~\cite{NCQ-scalling}
\begin{equation} \label{Toy:V2}
v_2/\mathcal{N} = a/(1+e^{-[(m_T-m_0)/\mathcal{N} -b]/c}) - d,
\end{equation}
where $\mathcal{N}=2$ is the number of constituent quarks in a pion, and $m_T$ and $m_0$ are its transverse mass and rest mass, respectively. The parameters ($a$, $b$, $c$ and $d$) are obtained by fitting Eq.~\ref{Toy:V2} to the experimental data~\cite{Wang:2016iov}. To add the resonance background, a fraction of primordial pions (33 $\pi^+$ and 33 $\pi^-$) are replaced with 33 $\pi^+$-$\pi^-$ pairs from $\rho$-meson decays, which is implemented with PYTHIA6~\cite{PYTHIA}. The $\rho$-meson $v_2$ is described by Eq.(\ref{Toy:V2}) with the corresponding $\rho$ masses. $v_3$ is fixed at $1/5$ of $v_2$ at any given $p_T$ for both primordial pions and $\rho$ resonances~\cite{STAR-rho-spct}.
The primordial-pion spectra  follow the Bose-Einstein distribution,
\begin{equation}
\frac{dN_{\pi^\pm}}{dm_T^2} \propto (e^{m_T/T_{\rm BE}}-1)^{-1},
\end{equation}
where $T_{\rm BE} = 212$ MeV is set to match the experimentally observed $\langle p_T \rangle$ of 400 MeV~\cite{STAR-pion-spct}. The $\rho$-resonance spectra obey 
\begin{equation}
\frac{dN_\rho}{dm_T^2} \propto \frac{e^{-(m_T-m_\rho)/T}}{T(m_\rho+T)},
\end{equation}
where $T = 317$ MeV is used to match its $\langle p_T \rangle$ of 830 MeV as observed in data~\cite{STAR-rho-spct}. Pseudorapidity (Rapidity) is uniformly distributed in the range of [-1, 1] for primordial pions ($\rho$ resonances). 

\subsection{EBE-AVFD}

The EBE-AVFD model~\cite{Shi:2017cpu,Jiang:2016wve,Shi:2019wzi} is a comprehensive simulation framework that dynamically describes  the CME in heavy-ion collisions. This state-of-the-art tool has been developed over the past few years as an important part of the  efforts within the Beam Energy Scan Theory (BEST) Collaboration, aiming to address the needs of the ongoing experimental program at RHIC collision energies.   Critical to the success of the CME search is a quantitative and realistic characterization of the CME signals as well as the relevant backgrounds. Accordingly,  EBE-AVFD  implements the dynamical CME transport for quark currents on top of the relativistically expanding viscous QGP fluid and properly models   major sources of background correlations such as LCC and resonance decays.  

More specifically, the EBE-AVFD framework starts with event-wise fluctuating initial conditions, and solves the evolution of chiral quark currents as linear perturbations in addition to the viscous bulk flow background provided by data-validated hydrodynamic simulation packages. The  LCC effect is incorporated in the freeze-out process, followed by the hadron cascade simulations. 
This is illustrated in the following Fig.~\ref{fig.avfd_flow_chart}.

\begin{figure}[!hbt]\centering
\includegraphics[width=0.8\textwidth]{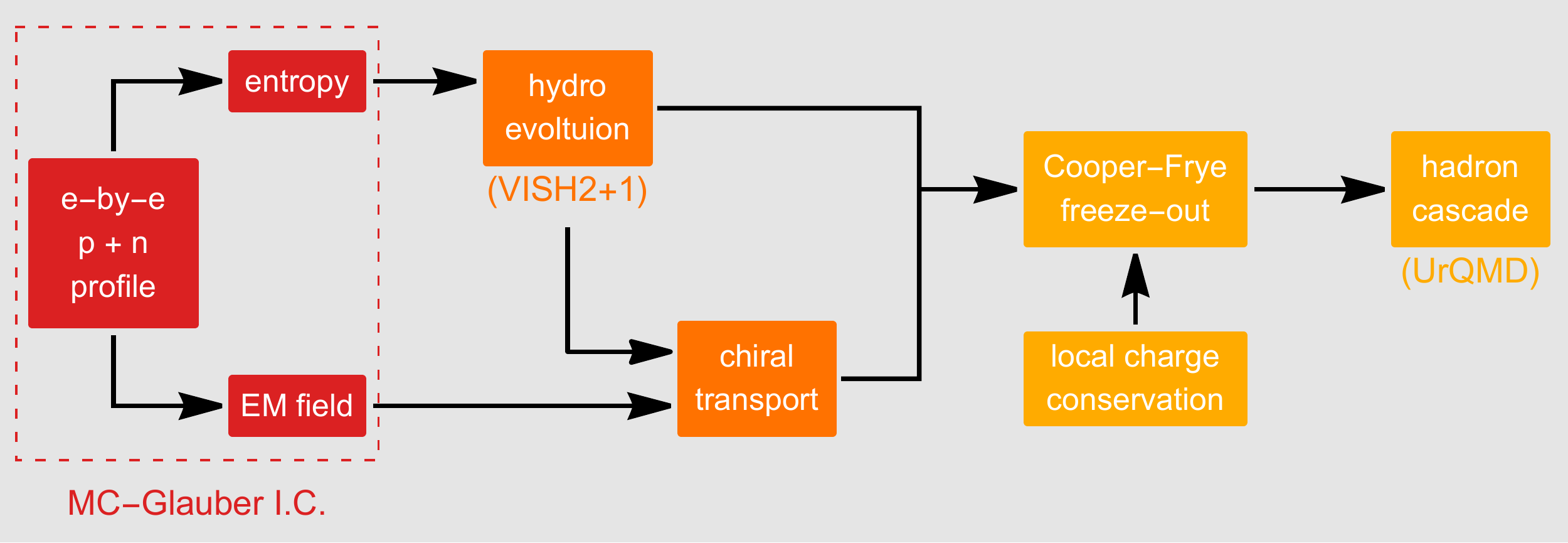}
\caption{Flow chart of the EBE-AVFD framework.\label{fig.avfd_flow_chart}}
\end{figure}

The fluctuating initial conditions for entropy density profiles are generated by the Monte-Carlo Glauber model, with switching time $\tau_0=0.6~\text{fm}/c$ and mixing parameter $\alpha_\text{glb} = 0.118$. The latter factorizes the contributions of the participant density ($n_\text{part}$) and binary collision density ($n_\text{coll}$) in the local entropy density, $s=\alpha_\text{glb}\, n_\text{coll} + (1-\alpha_\text{glb})n_\text{part}/2$. 
The initial axial charge density ($n_5$) is approximated in such a way that it is proportional to  the corresponding local entropy density with a constant ratio. This ratio parameter can be varied to sensitively control the strength of the CME transport. For example, one can set  $n_5/s$ to $0$, $0.1$ and $0.2$  in the simulations to represent scenarios of zero, modest and strong CME signals respectively. The initial electromagnetic field is computed according to the event-wise proton configuration in the Monte-Carlo Glauber initial conditions.

The hydrodynamic evolution is solved through two components. The bulk-matter collective flow is described by the VISH2+1 simulation package~\cite{Shen:2014vra}, with the lattice equation of state \texttt{s95p-v1.2}, shear-viscosity $\eta/s=0.08$, and freeze-out temperature $T_\text{fo}= 160~$MeV. Such hydrodynamic    simulations of bulk flow have been extensively tested and validated with relevant experimental data. The dynamical CME transport is described by anomalous hydrodynamic equations for the quark chiral currents on top of the bulk flow background, where the magnetic-field-induced CME currents lead to a charge separation in the fireball.   Additionally the conventional transport processes like diffusion and relaxation for the quark currents are consistently included, with the diffusion constant  chosen to be $\sigma=0.1\,T$ and relaxation time $\tau_r = 0.5/T$. More discussions of the hydrodynamics equations and relevant details can be found in Refs.~\cite{Shi:2017cpu,Jiang:2016wve,Shi:2019wzi}. 

After the hydrodynamic stage, hadrons are locally produced in all fluid cells on the freeze-out hypersurface, using the Cooper-Frye freeze-out formula 
\begin{eqnarray}\label{eq_cooperfrye}
E \frac{dN}{d^3p} (x^\mu, p^\mu) = \frac{g}{(2\pi)^3} \int_{\Sigma_{\rm fo}} p^\mu d^3\sigma_\mu f(x,p) \,\, .
\end{eqnarray} 
Here, the local distribution function automatically  includes the charge separation effect due to the CME as well as  non-equilibrium corrections. 
In the freeze-out process, the LCC effect is implemented by extending an earlier method from Ref.~\cite{Schenke:2019ruo}. The approach in Ref.~\cite{Schenke:2019ruo} chooses to produce all charged hadron-antihadron pairs at the same fluid cell, while their momenta are sampled independently in the local rest frame of the fluid cell. This treatment implicitly assumes the charge-correlation length to be smaller than the cell size, and hence provides an upper limit for the correlations between opposite-sign pairs. In the EBE-AVFD package, the aforementioned procedure is generalized and improved to mimic more realistically the impact of a finite charge-correlation length: a new parameter $P_\text{LCC}$ is introduced to characterize the fraction of charged hadrons that are sampled in positive-negative pairs in the same way as in Ref.~\cite{Schenke:2019ruo}, while the rest of the hadrons are sampled independently. Varying the parameter $P_\text{LCC}$ between 0 and 1 would tune the LCC contributions from none to its maximum. 
Finally, all the hadrons produced from the freeze-out hypersurface are further subject to hadron cascades through the UrQMD simulations~\cite{Bleicher:1999xi}, which  account for various hadron resonance decay processes and automatically include their contributions to the charge-dependent correlations. The tuning of the EBE-AVFD calculations to the experimental measurements of $\Delta\delta$  and $\Delta\gamma_{112}$ in Au+Au collisions suggests that an optimal value of   $P_\text{LCC}$ is around $1/3$, and that roughly half of the background correlations come  from LCC and the other half from resonance decays. 

\section{Core-Component Comparisons}
\label{Sec:kernel}
We will exploit the toy model and the EBE-AVFD model to simulate the core components of
the experimental observables introduced in Sec.~\ref{Sec.II}:
$\Delta \gamma_{112}$  for the $\gamma$ correlator, $\Delta_{R2}$ for the $R$ correlator, and $\Delta_{\rm SBF}$ for the signed balance functions.
Our objective is to examine the responses of the core components to the CME signal and the background, and to verify the relations between these methods (Eqs.~\ref{eq:relation1} and \ref{eq:relation3}).
For a fair comparison with other observables, the momentum weighting is not applied in the $\Delta_{\rm SBF}$ results.
For simplicity, the true reaction plane is used in all the simulations in this section. The particles of interest are selected with $|\eta|<1$ and $0.2 < p_T < 2$ GeV/$c$.

\subsection{Toy-model Results}

\begin{figure}[tbp] 
\includegraphics[width=6.2cm]{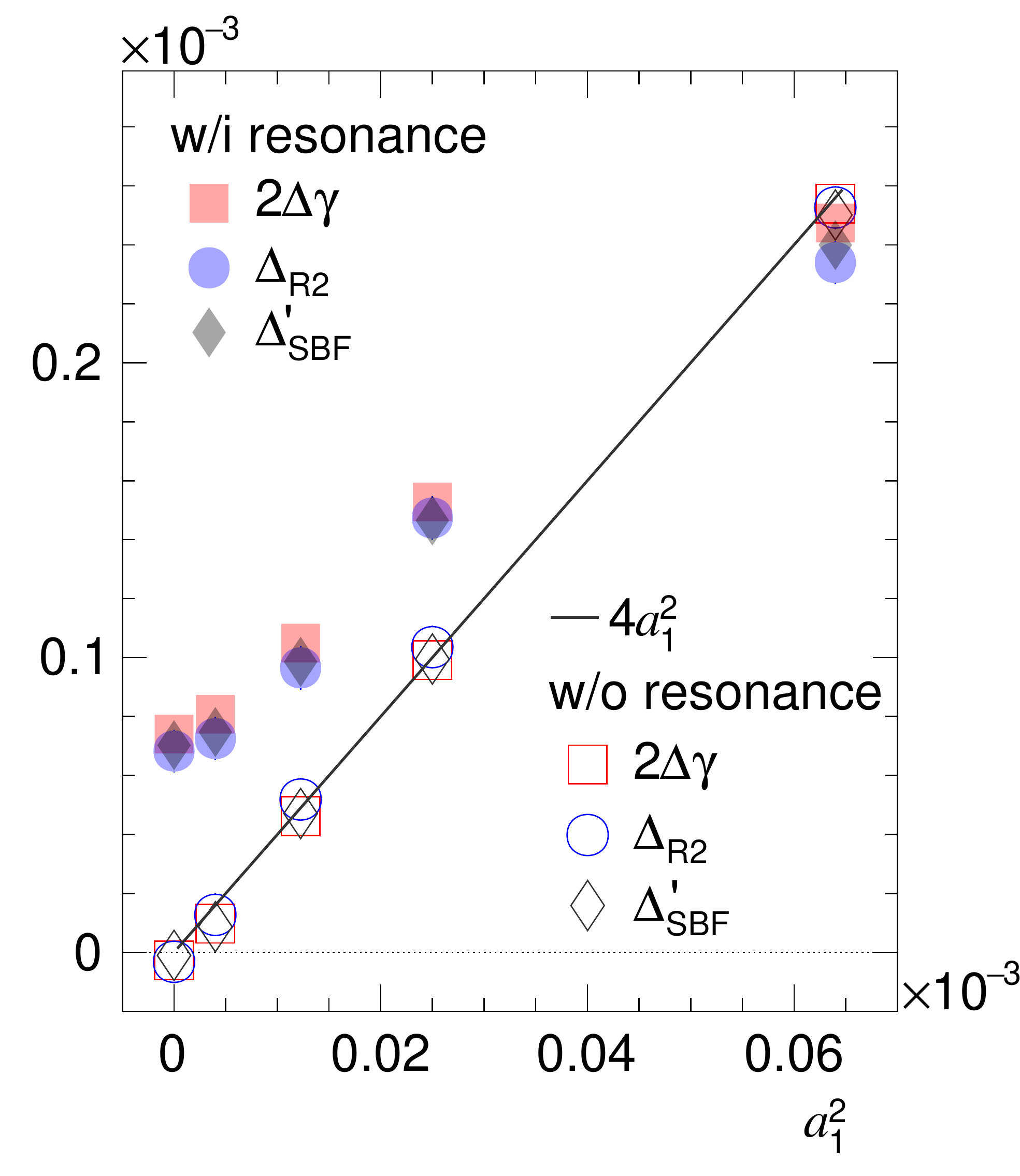}
\caption{The toy-model simulations of $2\Delta\gamma_{112}$, $\Delta_{R2}$ and $\Delta'_{\rm SBF}\equiv(\frac{\pi^4}{64M^2}\Delta_{\rm SBF}+\frac{8}{3}v_2\Delta\delta)$ as function of the input $a_1^2$. The open markers represent the  pure-signal scenario without resonances, and the solid markers denote the scenario with resonance decays. In comparison, the linear function of $4a^2_1$ is also added.} \label{fig:R_psi2}
\end{figure}

We put the three core components on the same footing,
by plotting $2\Delta\gamma_{112}$, $\Delta_{R2}$ and $\Delta'_{\rm SBF}\equiv(\frac{\pi^4}{64M^2}\Delta_{\rm SBF}+\frac{8}{3}v_2\Delta\delta)$ as function of the input $a_1^2$ in Fig.~\ref{fig:R_psi2} for two scenarios: with and without resonance decays (LCC is not implemented in the toy model).
In the background-free case without decays, the three observables render very similar results (open markers), all falling on the linear function of $4a_1^2$. Therefore, all the three methods are sensitive to the same amount of the CME signal. 
When resonance decays are turned on with finite elliptic flow, sizeable background effects appear besides the pure-signal contributions for all the three approaches (solid markers), more prominent at smaller input $a_1^2$ values. 
Note that each final observable can be roughly regarded as a weighted average of the correlations due to the CME, the resonance background and the cross terms.
At unrealistically large $a_1^2$ values, the CME contribution in an observable could be diluted by the resonance contribution, since the latter becomes smaller than the former. The three core components exhibit similar responses to the backgrounds due to flowing resonances in this toy model.
In this scenario, there are some subtle differences between the results from these three approaches, probably because of some higher-order effects omitted in the derivation of Eqs.~\ref{eq:relation1} and \ref{eq:relation3}.
Although the background contributions depend on  spectra and particularly  elliptic flow of resonances~\cite{Feng:2018chm,Schlichting:2010qia,Pratt:2010zn}, to the first order, we expect the three observables to have similar responses to resonance decays for a wide range of spectra or elliptic flow.
A recent study~\cite{PhysRevC.103.034912} has also found that the $R$ correlator and the $\Delta\gamma_{112}$ correlator have similar sensitivities to the CME signal and the background.

\subsection{EBE-AVFD Results}

\begin{figure}[tbp]
\vspace*{-0.01in}
\subfigure{\includegraphics[width=6.2cm]{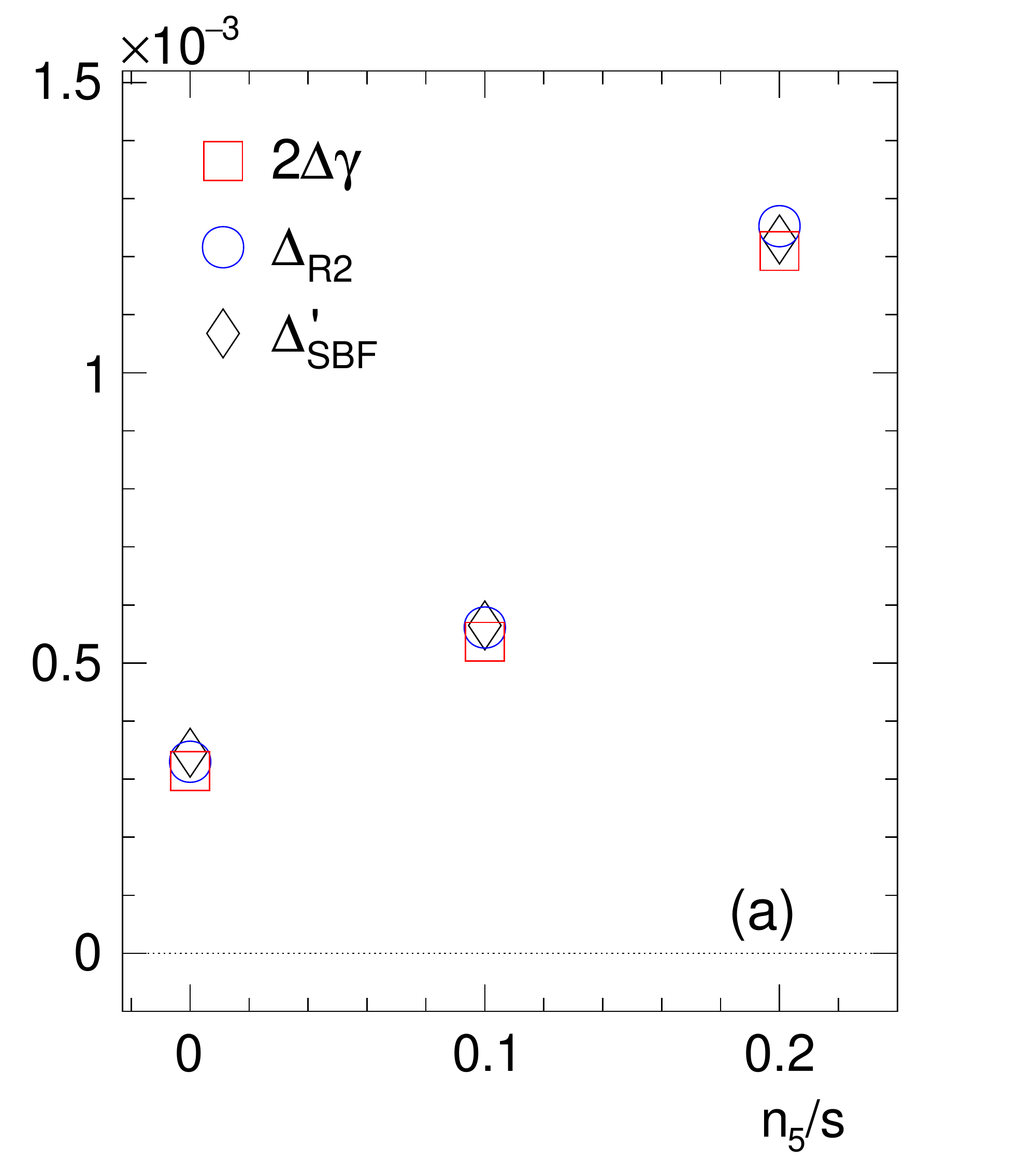}}
\subfigure{\includegraphics[width=6.2cm]{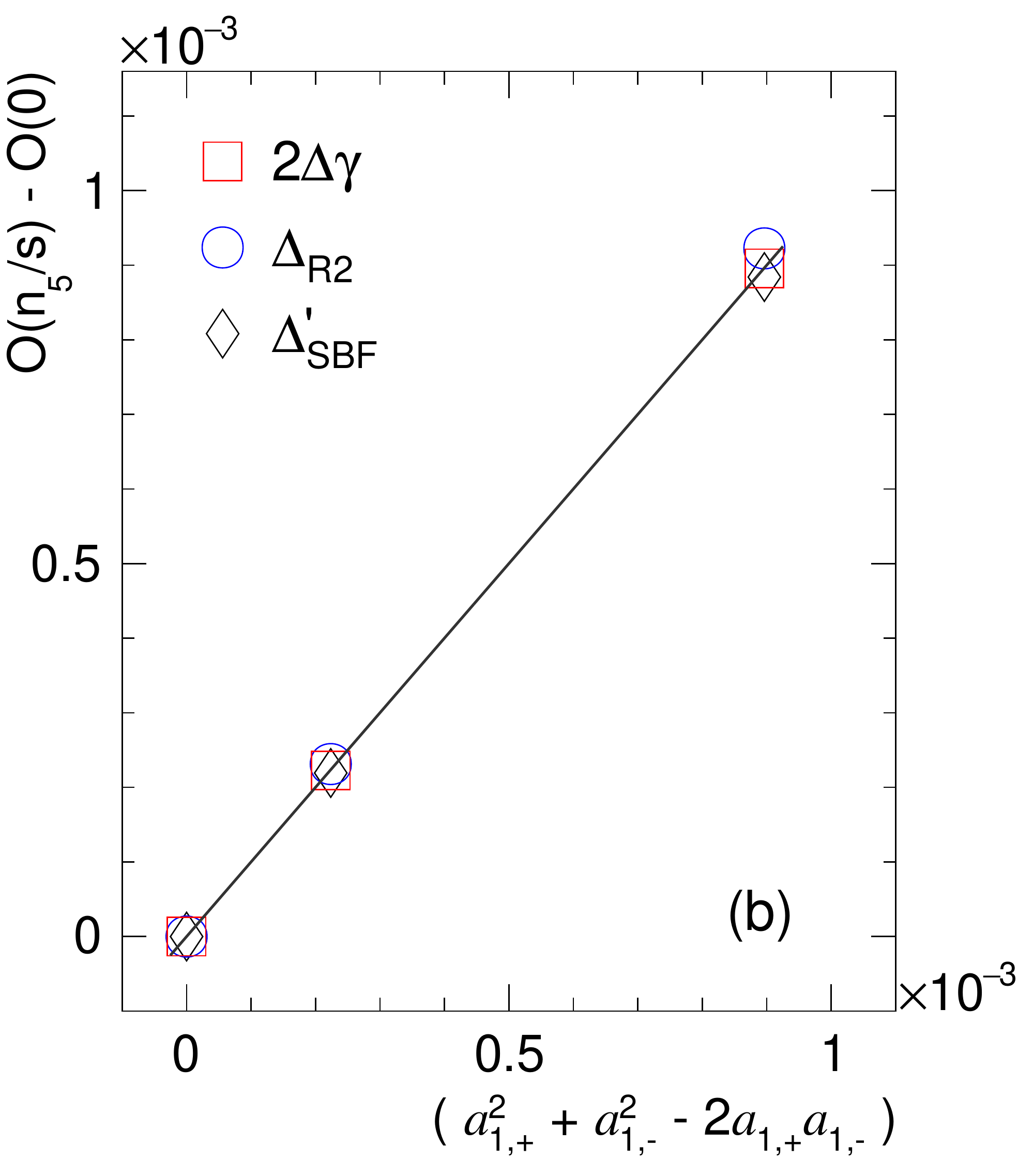}}
\captionof{figure}{(a) The EBE-AVFD simulations of $2\Delta\gamma_{112}$, $\Delta_{R2}$ and $\Delta'_{\rm SBF}\equiv(\frac{\pi^4}{64M^2}\Delta_{\rm SBF}+\frac{8}{3}v_2\Delta\delta)$ as function of $n_5/s$ in 30-40\% Au+Au collisions at 200 GeV. (b) The same results with the subtraction of the pure-background case vs $(a_{1,+}^2 + a_{1,-}^2 - 2a_{1,+}a_{1,-})$. In comparison, a linear function of $y=x$ is drawn to verify the relation in Eq.~\ref{eq:Superposition}.} \label{fig:AVFD_delta}
\end{figure}

The EBE-AVFD model implements the CME and the backgrounds in a more realistic way. 
In the following simulations, we generate the EBE-AVFD events of 30-40\% Au+Au collisions at $\sqrt{s_{\rm NN}} = 200$ GeV, with $n_{5}/s$ = 0, 0.1 and 0.2.
The background effects almost remain the same, whereas the CME signal is varied according to the input $n_{5}/s$.
Figure~\ref{fig:AVFD_delta}(a) presents the corresponding calculations of $2\Delta\gamma_{112}$, $\Delta_{R2}$ and $\Delta'_{\rm SBF}$ as function of $n_{5}/s$.
The three methods yield very similar results at each  input $n_{5}/s$ value, supporting the relations expressed in 
Eqs.~\ref{eq:relation1} and \ref{eq:relation3}.

With the known reaction plane angle in each EBE-AVFD event, we can readily calculate $a_{1,\pm}$, and check if this CME contribution explains the difference between the cases with different $n_{5}/s$ values. $a_{1,\pm}$ is consistent with zero for $n_{5}/s=0$, and $a_{1,+}$ and $a_{1,-}$
are finite with opposite signs for finite $n_{5}/s$ values. Note that $a_{1,+}$ and $a_{1,-}$ do not necessarily have the same magnitude, because the collision system always bears extra positive charges.
Based on the expansion of the $\gamma_{112}$ correlator in Eq.~\ref{eq:ThreePoint} and the equivalence between the three observables, we expect the following 
equation for any of these observables, $O(n_{5}/s)$, 
\begin{equation}
O(n_{5}/s) - O(0) =  a_{1,+}^2 + a_{1,-}^2 - 2a_{1,+}a_{1,-}.
\label{eq:Superposition}
\end{equation}
Figure~\ref{fig:AVFD_delta}(b) shows that the results for each  observable, after the subtraction of the pure-background case, fall on the straight line representing the relation in Eq.~\ref{eq:Superposition}. Thus, the EBE-AVFD calculations reveal the linear superposition of the CME signal and the background contribution in the experimental observables. 
This property is implicitly assumed by  most of the analysis techniques that attempt to separate the CME signal and the backgrounds, and it is now corroborated by the EBE-AVFD model. 

The core-component comparison using both the toy model and the EBE-AVFD model support the idea that to the first order, the three observables are equivalent to each other, with their very similar responses to the CME signal as well as the backgrounds.

\section{Sensitivity study for isobaric collisions}
\label{sectV}

The validity of Eq.~\ref{eq:Superposition} for the three experimental observables reassures the feasibility of disentangling the CME signal from the backgrounds with the isobaric collisions. To eliminate unintentional biases in the analyses of the isobaric-collision data, the STAR Collaboration has followed a blind-analysis procedure~\cite{STARBlinding}. One important step in this procedure requires the analysis codes from all analyzers to be frozen before the mass production of the isobaric-collision data. 
In this section, we shall apply these frozen codes
to the EBE-AVFD events of isobaric collisions,
to investigate the realistic sensitivity of each method to the CME signal. The particles of interest come from the acceptance of $|\eta|<1$, with $0.2 < p_T < 2$ GeV/$c$ for the $\gamma$ correlator and the signed balance functions,
or with $0.35 < p_T < 2$ GeV/$c$ for the $R$ correlator, as implemented in the frozen codes.
Owing to different kinematic cuts used in different methods, the sensitivity study in this manner cannot guarantee apples-to-apples comparisons as was assured in the core-component comparisons in Sec~\ref{Sec:kernel}. Nevertheless, we will obtain a reliable benchmark for interpreting experimental data. 

For each of the two isobaric collision systems, Ru+Ru and Zr+Zr at $\sqrt{s_{\rm NN}} = 200$ GeV, four cases of the EBE-AVFD events have been generated, with $n_5/s = 0,\,  0.05,\, 0.1, \, \mathrm{and} \, 0.2$, respectively. The centrality selection for all the cases focuses on 30-40\% central collisions, where the potential CME signal is  relatively easy to detect owing to good event plane resolutions. 
200 million events are produced for each case of $n_5/s = 0$ and $n_5/s = 0.2$, and 400 million events for each of the other cases. To mimic the detection performance of the STAR Time Projection Chamber, the simulated particles in the EBE-AVFD events are randomly rejected according to a transverse-momentum dependent tracking efficiency.

Table~\ref{tab:Observeda1} lists $a_{1,\pm}$  calculated by EBE-AVFD at different $n_5/s$ values. 
Basically 
$a_{1,\pm}$ displays a linear function of $n_5/s$.  $a_{1,+}$ is about 4\% larger than $a_{1,-}$ for all the cases, reflecting the charge asymmetry in the collision systems.
\begin{center}
\begin{table}[b]
\centering
\caption{The $a_{1,\pm}$ values calculated for the EBE-AVFD events of 30-40\% isobaric collisions at $\sqrt{s_{\rm NN}} = 200$ GeV.}
\resizebox{0.36\textwidth}{!}{
\begin{tabular}{c|cc|cc}
\toprule
\multirow{2}{*}{ $n_{5}/s$}       &  \multicolumn{2}{c}{  $a_{1,+}$ (\%)}      & \multicolumn{2}{|c}{ $a_{1,-}$ (\%)  }  \\   \cline{2-5}
~ & Ru+Ru & Zr+Zr  & Ru+Ru & Zr+Zr  \\ 
\hline
0   &  0 & 0 &  0 & 0\\
0.05 &  0.37 &  0.35  &  0.35  & 0.33 \\
0.10 &  0.74 &  0.69  &  0.71  & 0.66 \\
0.20 &  1.48  &  1.38  &  1.42  & 1.32 \\
\bottomrule
\end{tabular}
}
\label{tab:Observeda1}
\end{table}
\end{center}
Meanwhile, $a_{1,\pm}$ is roughly 7\% larger in Ru+Ru than in Zr+Zr collisions, which meets the expectation from the difference in the corresponding magnetic fields~\cite{isobar1}.
For the  experimental observables under study, 
the background levels are almost the same for the two isobaric systems, and the ratio of the  Ru+Ru measurement to the Zr+Zr measurement is expected to be larger than unity, in the presence of a positive CME signal.
Hence, the sensitivity of each method can be defined with the statistical significance of the deviation of this ratio from unity. 

\begin{figure}[tb]\centering
\subfigure{\includegraphics[width=6.2cm]{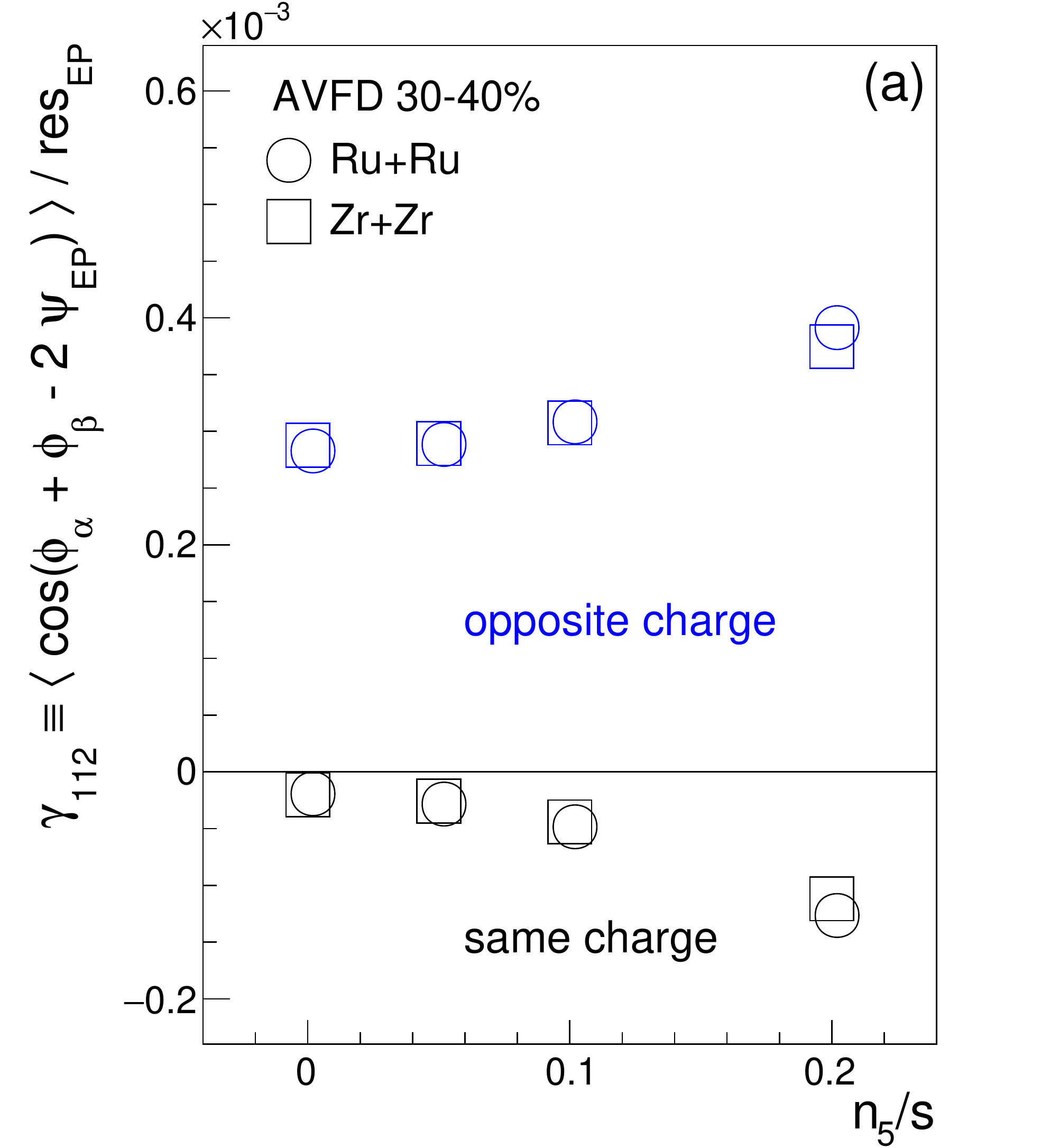}}
\subfigure{\includegraphics[width=6.2cm]{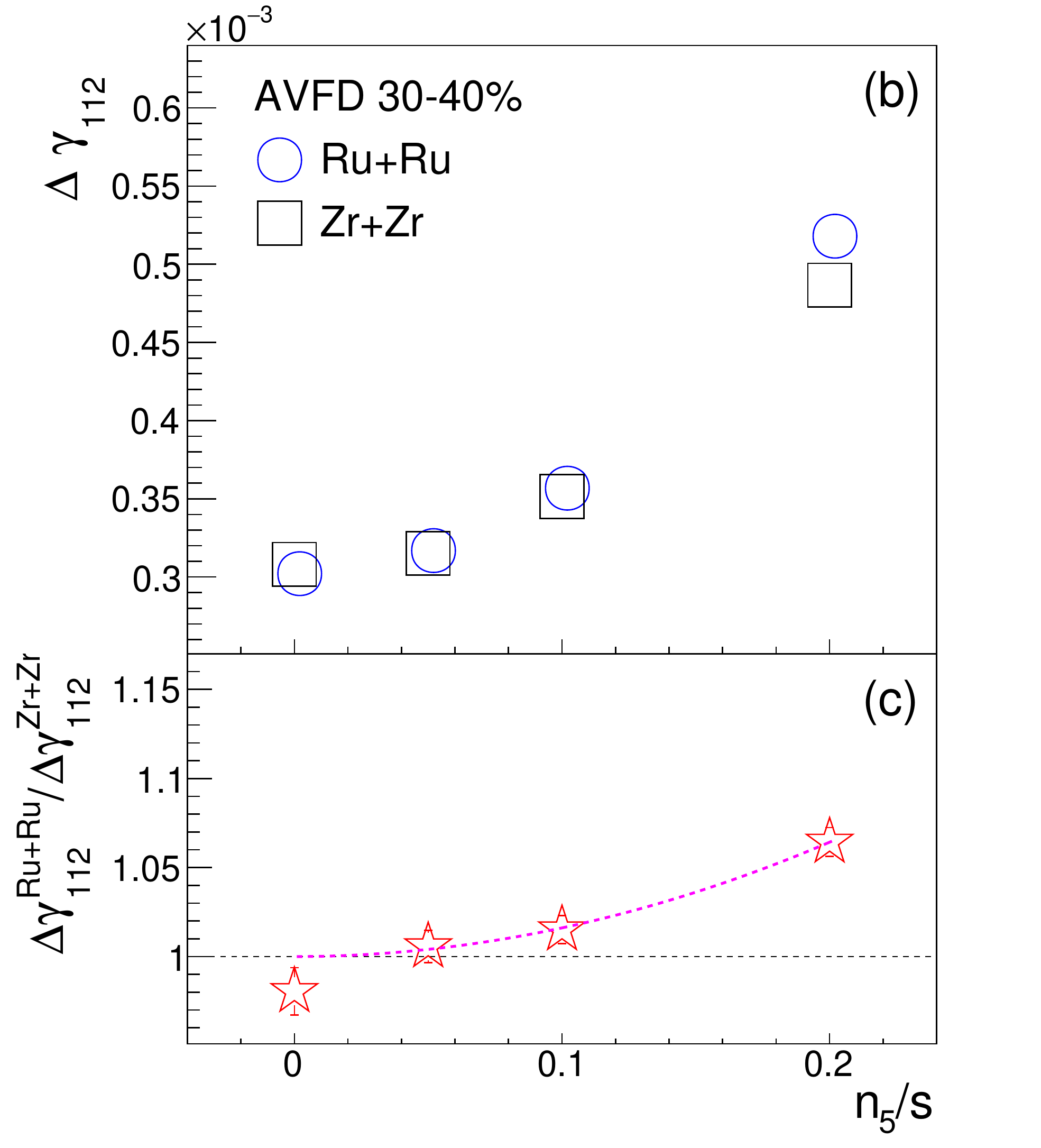}}
\caption{EBE-AVFD calculations of  $\gamma_{112}^{\rm OS(SS)}$ (a) and $\Delta \gamma_{112}$ (b) as functions of $n_{5}/s$ for 30-40\% isobaric collisions at $\sqrt{s_{\rm NN}} = 200$ GeV, together with the ratio of   $\Delta \gamma_{112}$ (c) between Ru+Ru and Zr+Zr.
In panel (c), the $2^{\rm nd}$-order-polynomial fit function illustrates the rising trend starting from (0, 1). 
}
\label{gamma_isobar}
\end{figure}

Figure~\ref{gamma_isobar} presents the EBE-AVFD calculations of  $\gamma_{112}^{\rm OS(SS)}$ (a) and $\Delta \gamma_{112}$ (b) as functions of $n_{5}/s$ for 30-40\% isobaric collisions at $\sqrt{s_{\rm NN}} = 200$ GeV. The ratios of $\Delta \gamma_{112}$  between Ru+Ru and Zr+Zr is delineated in panels (c).
The $2^{\rm nd}$-order event plane is reconstructed from the same kinematic region as the particles of interest, and the observed $\gamma$ correlators and $v_2$ (to be shown later) have been corrected with the corresponding event plane resolution.
At each $n_5/s$ value, $\gamma_{112}^{\rm OS}$ remains positive and $\gamma_{112}^{\rm SS}$ stays negative, both with larger magnitudes at higher $n_5/s$. Although the CME expects $\gamma_{112}^{\rm OS}$ and $\gamma_{112}^{\rm SS}$ to be symmetric around zero, there exist some charge-independent backgrounds such as momentum conservation and elliptic flow that shift both $\gamma_{112}^{\rm OS}$ and $\gamma_{112}^{\rm SS}$ up or down~\cite{STAR3}.
Therefore, we shall focus on 
$\Delta\gamma_{112}$,
which shows a finite background contribution at $n_5/s=0$ and increases with the CME signal. The difference between Ru+Ru and Zr+Zr is better viewed with the ratio of $\Delta\gamma_{112}^{\rm Ru+Ru}/\Delta\gamma_{112}^{\rm Zr+Zr}$. This ratio is consistent with unity at $n_5/s=0$, and increases quadratically
with $n_5/s$ as demonstrated by the $2^{\rm nd}$-order-polynomial fit function that passes (0, 1) (dashed line). 
The quadratically-increasing trend is expected, because this ratio is a linear function of the CME signal fraction in $\Delta\gamma_{112}$ in a two-component
perturbative framework~\cite{isobar1}, and the latter is proportional to $(n_5/s)^2$ or $a_1^2$ as shown in Fig.~\ref{fig:AVFD_delta}.

\begin{figure}
\subfigure{\includegraphics[width=6.2cm]{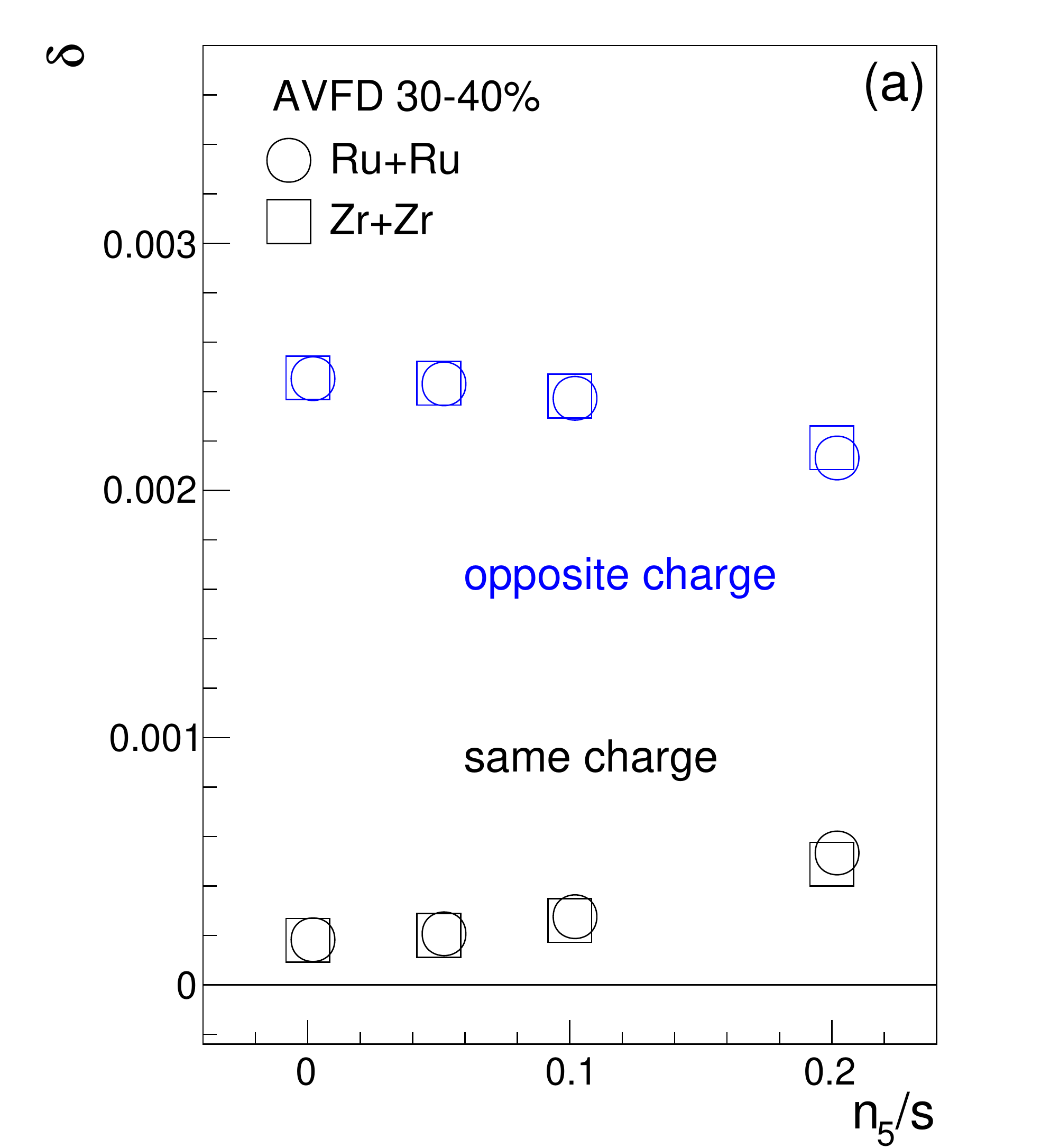}}
\subfigure{\includegraphics[width=6.2cm]{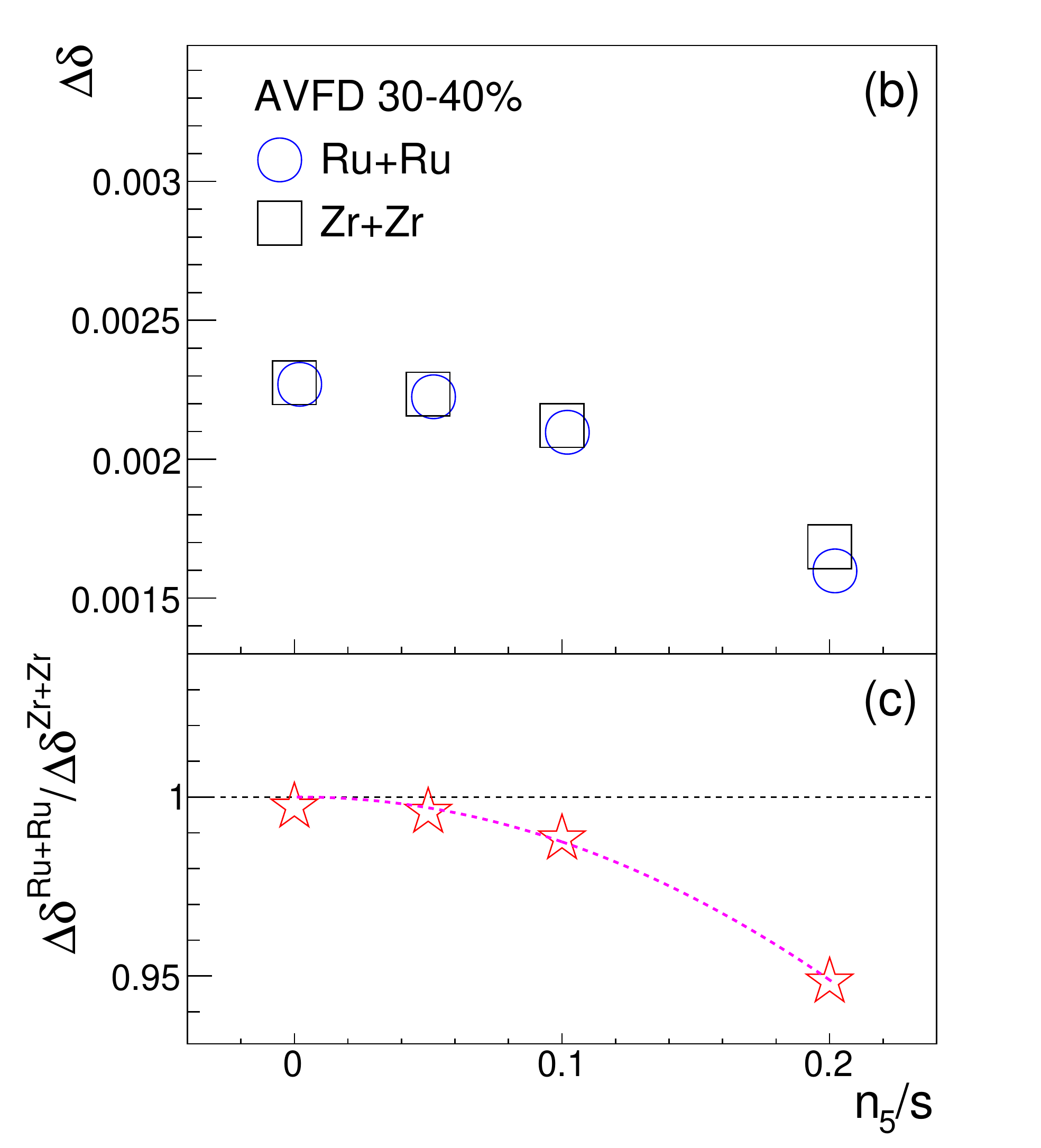}}
\caption{EBE-AVFD calculations of  $\delta^{\rm OS(SS)}$ (a) and $\Delta \delta$ (b) as functions of $n_{5}/s$ for 30-40\% isobaric collisions at $\sqrt{s_{\rm NN}} = 200$ GeV, together with the ratios of $\Delta \delta$ (c) between Ru+Ru and Zr+Zr.
In panels (c) , the $2^{\rm nd}$-order-polynomial fit function is added to demonstrate the rising trend starting from (0, 1). 
}
\label{delta_isobar}
\end{figure}

Potentially the $\delta$ correlator could also contain the CME signal as Eq.~\ref{eq:delta} suggests, which should be manifested in the $\Delta\delta$ ratio between Ru+Ru and Zr+Zr~\cite{Shi:2019wzi}. 
Figure~\ref{delta_isobar} depicts 
the EBE-AVFD results of  $\delta^{\rm OS(SS)}$ (a) and $\Delta \delta$ (b) as functions of $n_{5}/s$ for 30-40\% isobaric collisions at $\sqrt{s_{\rm NN}} = 200$ GeV. The ratio of $\Delta \delta$  between Ru+Ru and Zr+Zr is shown in panel (c). For all the cases, $\delta^{\rm OS}$ is above $\delta^{\rm SS}$, leading to a positive $\Delta\delta$, because of larger background contributions to $\delta^{\rm OS}$ than $\delta^{\rm SS}$.  It is argued that although overshadowed by the background contributions, the CME signal in $\Delta\delta$ could be even larger than that in $\Delta\gamma_{112}$ measured with respect to the participant plane~\cite{Shi:2019wzi}. This is because the magnetic field difference between the two isobaric systems is maximal with respect to the reaction plane, and is reduced when measured otherwise. The EBE-AVFD simulations indeed support this idea: at each $n_5/s$ value in Fig.~\ref{delta_isobar}(c), the deviation of the $\Delta\delta$ ratio
from unity is  at the similar level as that of the $\Delta\gamma_{112}$ ratio.
Since $\Delta\delta$ is typically  larger than $\Delta\gamma$ by an order of magnitude, this similar deviation in their respective ratios indicates a significantly larger CME effect in $\Delta\delta$ than in $\Delta\gamma$.
A $2^{\rm nd}$-order polynomial fit function is added to guide the eye.
In view of the smaller relative statistical uncertainties of $\Delta\delta$ than those of $\Delta\gamma_{112}$, the former may yield even better significance levels of the CME signal than the latter in the isobaric-collision data, provided equal background contributions to $\Delta\delta$ in the two systems.
The caveat is that the two-particle correlation background to $\Delta\delta$ is significantly larger than that to $\Delta\gamma$, so any difference in the background between the two isobaric systems would have a stronger impact on $\Delta\delta$.
However likely or unlikely this scenario is to occur, we keep these results for completeness and leave them to the test with the real data.
Between Ru+Ru and Zr+Zr, the difference in the background contributions to $\Delta\gamma_{112}$ may be small, but could still be finite owing to the possibly different $v_2$ values. 
The normalization of $\Delta\gamma_{112}$ by $v_2$ would be a more robust variable for the CME search.

\begin{figure}[!hbt]
\begin{center}
\hspace{-1.0cm}
\subfigure{\includegraphics[width=6.2cm]{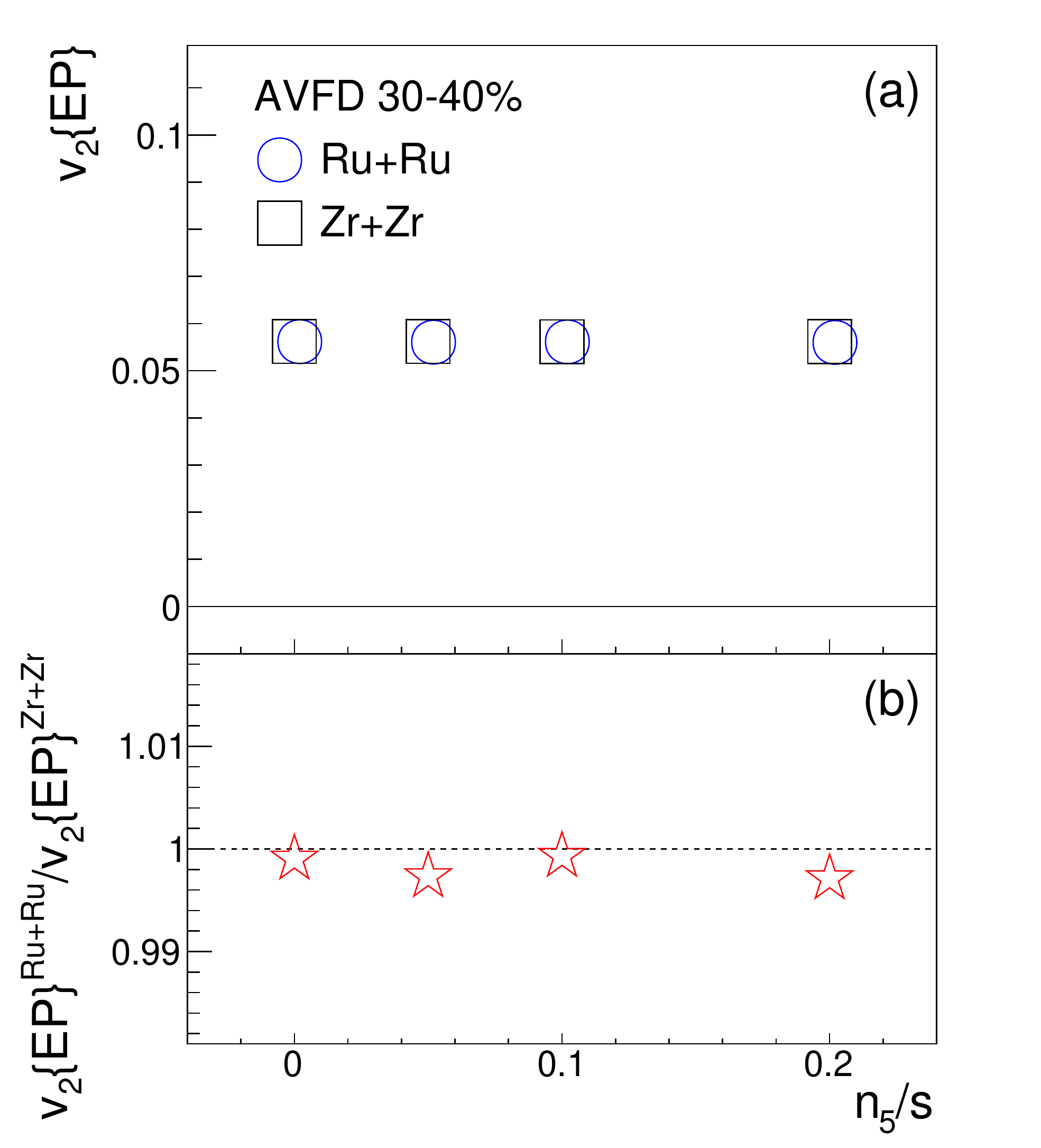}}
\subfigure{\includegraphics[width=6.2cm]{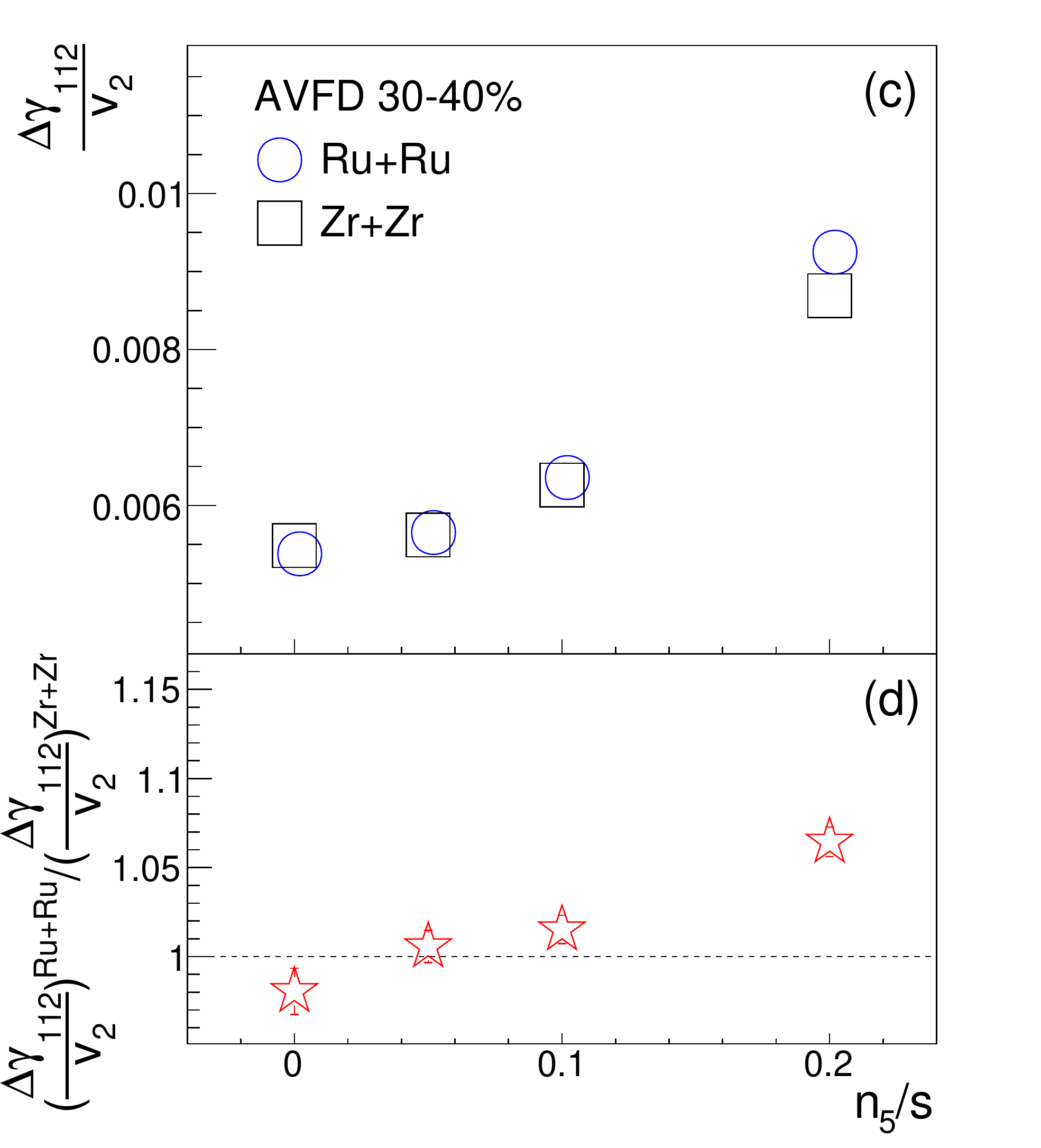}}
\subfigure{\includegraphics[width=6.2cm]{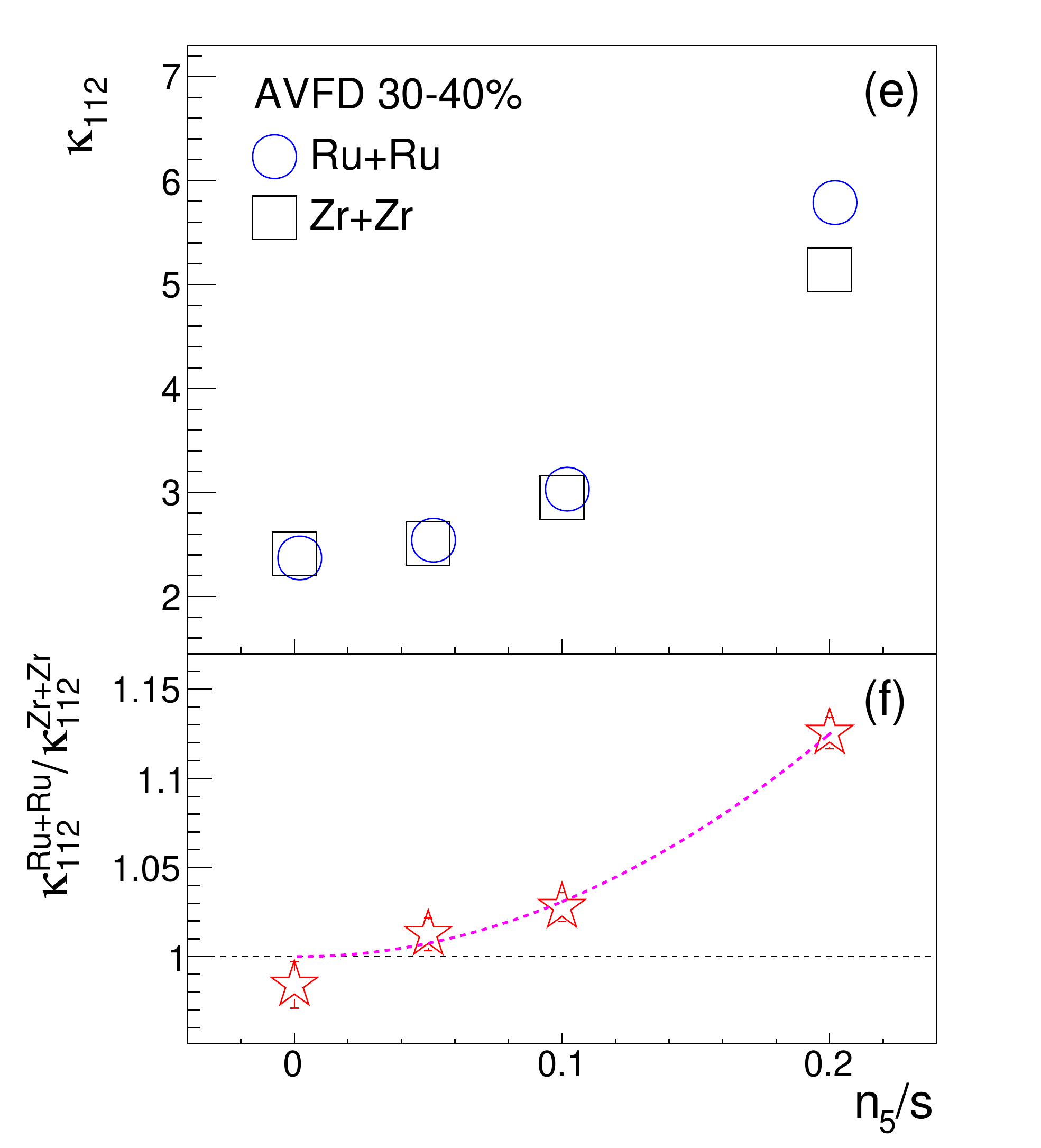}}
\end{center}
\caption{EBE-AVFD calculations of  $v_2$ (a), $\Delta\gamma_{112}/v_2$ (c) and $\kappa_{112}$ (e) as functions of $n_{5}/s$ for 30-40\% isobaric collisions at $\sqrt{s_{\rm NN}} = 200$ GeV, together with the ratios of $v_2$ (b), $\Delta\gamma_{112}/v_2$ (d) and $\kappa_{112}$ (f) between Ru+Ru and Zr+Zr.
In panels (f), the $2^{\rm nd}$-order-polynomial fit function illustrates the rising trend starting from (0, 1). }
\label{kappa_isobar}
\end{figure}
The EBE-AVFD simulations of $v_2$ (a) and $\Delta\gamma_{112}/v_2$ (c) are presented in Fig.~\ref{kappa_isobar} as functions of $n_{5}/s$ for 30-40\% isobaric collisions at $\sqrt{s_{\rm NN}} = 200$ GeV,
with the corresponding $v_2$ ratio (b) and $\Delta\gamma_{112}/v_2$ ratio (d) between Ru+Ru and Zr+Zr. 
The $v_2$ values are very close to each other between Ru+Ru and Zr+Zr collisions, with the relative difference at the level of $0.1\%$.
Because of this, the Ru+Ru/Zr+Zr ratio of $\Delta\gamma_{112}/v_2$ in Fig.~\ref{kappa_isobar}(d) is practically identical to that of $\Delta\gamma$ in Fig.~\ref{gamma_isobar}(c).
Besides the possible $v_2$ difference, the two-particle correlation strengths (which is part of the CME background) could also differ from Ru+Ru to Zr+Zr collisions.
As we mentioned above, although $\Delta\delta$ is also sensitive to the CME,  it is overwhelmed by background correlations and thus may be used as an approximate gauge for the two-particle correlation strength. The additional normalization of  $\Delta\gamma_{112}/v_2$ by $\Delta\delta$ in $\kappa_{112}$ could, therefore, further suppress this difference, and enhance the sensitivity to the CME signal.
There are two potential scenarios where  $\kappa_{112}$ 
has the advantage
over $\Delta\gamma_{112}$.
First, in reality the $\Delta\delta$ ratio may disfavor the CME interpretation, and in that case, the $\kappa_{112}$ ratio is less prone to a faked CME signal than the $\gamma_{112}$ ratio. Second, $\Delta\delta$ may also contain the CME signal as shown in Fig.~\ref{delta_isobar}(c), while the relative statistical uncertainty of $\Delta\delta$ is smaller than that of $\Delta\gamma_{112}$.
In that case, $\kappa_{112}^{\rm Ru+Ru}/\kappa_{112}^{\rm Zr+Zr}$ could yield a larger ratio than $\Delta\gamma_{112}^{\rm Ru+Ru}/\Delta\gamma_{112}^{\rm Zr+Zr}$, with a similar relative statistical uncertainty,
which means a better significance of the CME signal. Indeed, panel (f) also shows a quadratically increasing trend, with larger magnitudes than panel (d) at finite $n_5/s$ values, and the calculated significance values for $\kappa_{112}^{\rm Ru+Ru}/\kappa_{112}^{\rm Zr+Zr}$ roughly double those for $\Delta\gamma_{112}^{\rm Ru+Ru}/\Delta\gamma_{112}^{\rm Zr+Zr}$, as documented later in Table~\ref{tab:significance}.
In general, the EBE-AVFD simulations show good responses to the signal for
$\Delta\gamma_{112}^\mathrm{Ru+Ru} / \Delta\gamma_{112}^\mathrm{Zr+Zr}$, $\Delta\delta^\mathrm{Ru+Ru} / \Delta\delta^\mathrm{Zr+Zr}$ and $\kappa_{112}^\mathrm{Ru+Ru} / \kappa_{112}^\mathrm{Zr+Zr}$, and these promising features await verification by the isobaric-collision data.

\begin{figure}[!hbt]
\hspace{-1.2cm}
\centering
\subfigure{\includegraphics[width=6.2cm]{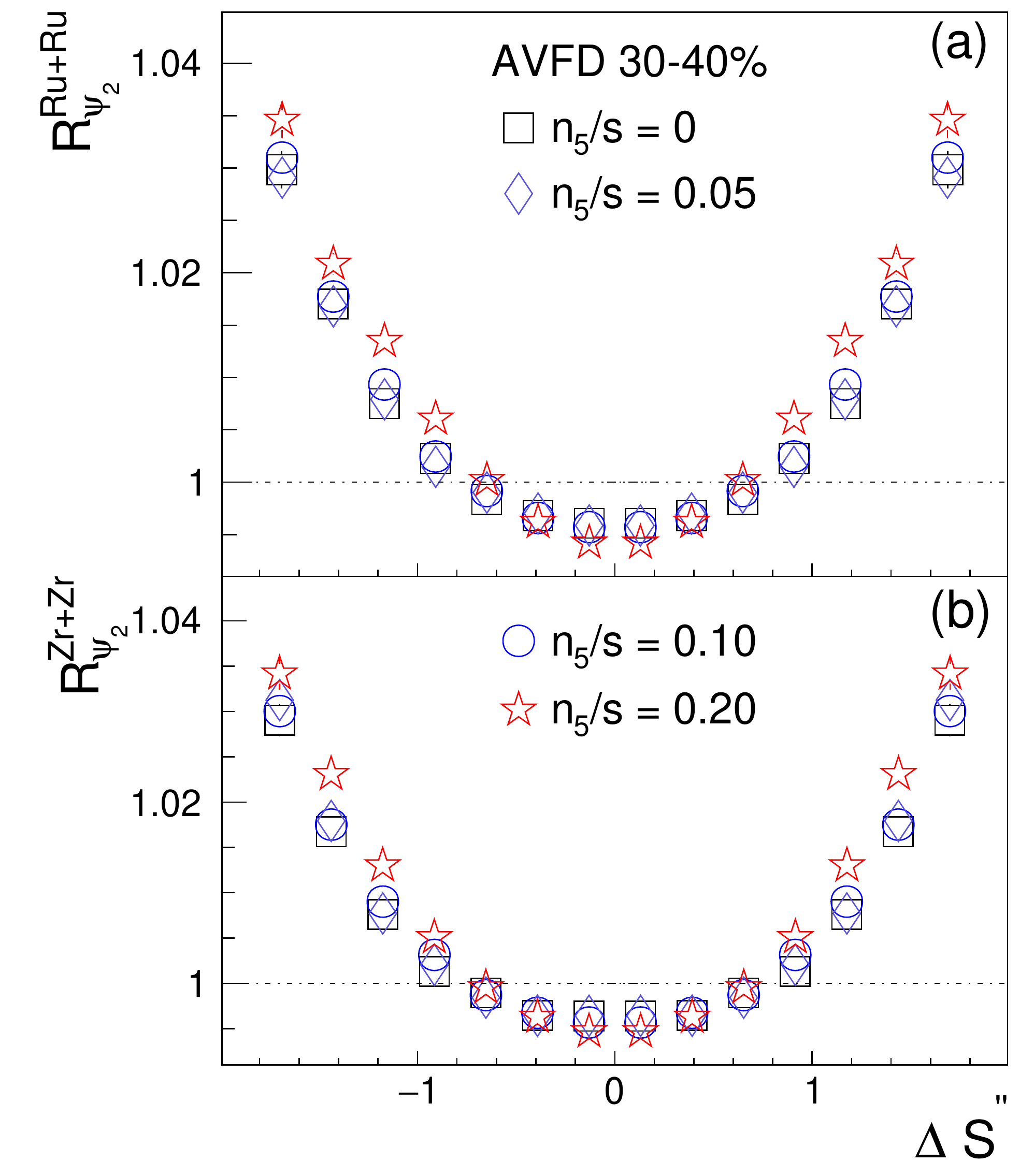}}
\subfigure{\includegraphics[width=6.2cm]{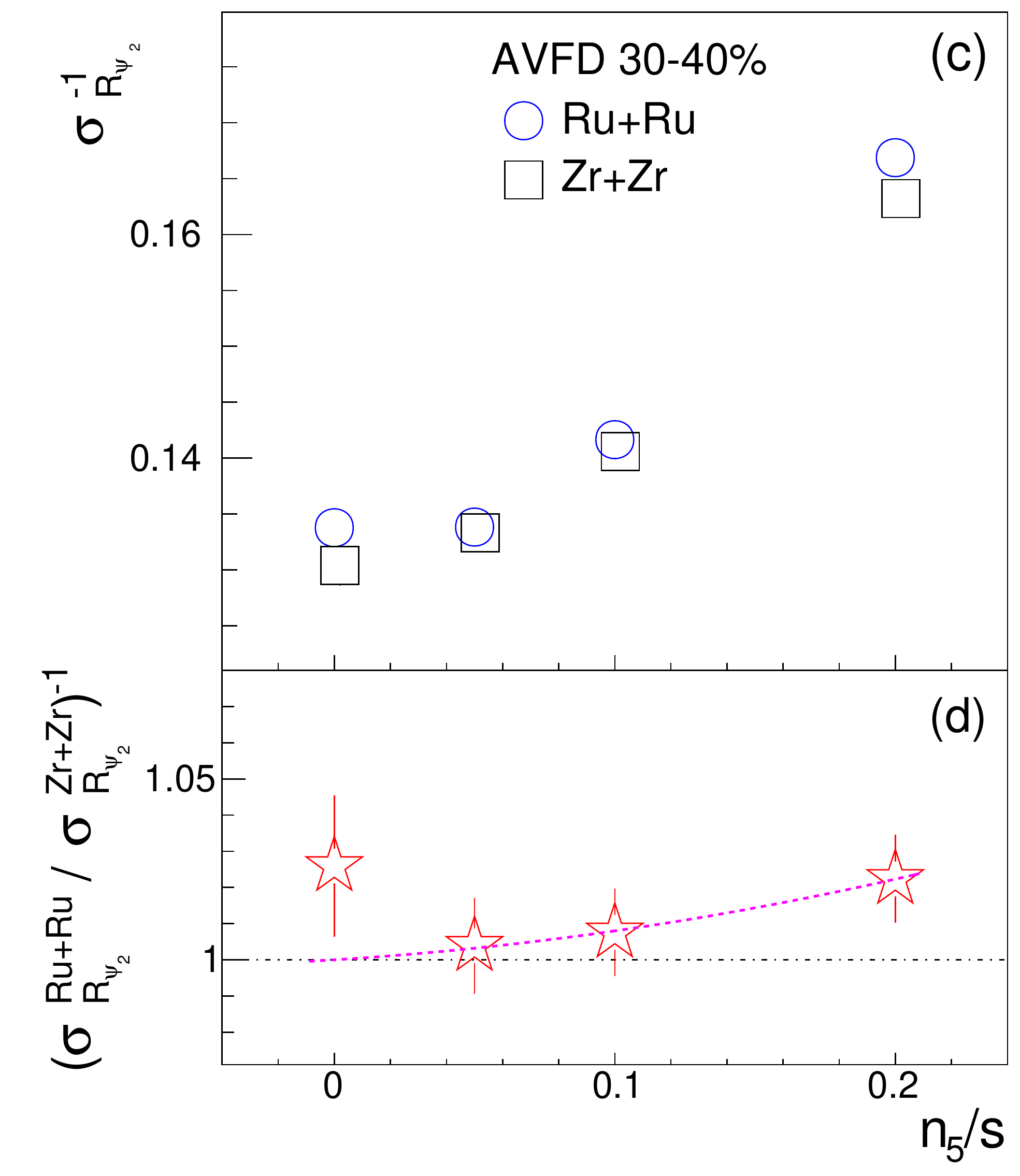}}
\caption{Distributions of $R(\Delta S_2^{''})$ from EBE-AVFD events of 30-40\% Ru+Ru (a) and Zr+Zr (b) at 200 GeV with different $n_{5}/s$ inputs. 
Panel (c) lists  $\sigma^{-1}_{R2}$ vs $n_{5}/s$, extracted from panels (a) and (b), and the $\sigma^{-1}_{R2}$ ratios between Ru+Ru and Zr+Zr are shown in panel (d), where the $2^{\rm nd}$-order-polynomial fit function shows the rising trend starting from (0, 1).
}
\hspace{1.43cm}\label{fig:R_isobar}
\end{figure}

A similar frozen-code analysis is performed for the $R(\Delta S_2)$ correlator, and the results are presented in Figure~\ref{fig:R_isobar}.
In order to minimize the influence of the particle number fluctuations, the $R(\Delta S_2)$ distribution is converted into the $R(\Delta S_2^{'})$ distribution by dividing 
the horizontal axis by the RMS of the $N(\Delta S_{2,\rm shuffled})$ distribution, {\em i.e.}, $\Delta S_2^{'} = \Delta S_2/\sqrt{\langle(\Delta S_{2,\rm shuffled})^2\rangle}$. 
Then $\Delta S_2^{'}$ is further modified to correct  for the event plane resolution, {\em i.e.}, 
$\Delta S^{''}= \Delta S^{'}/\mathrm{\delta_{Res}}$, where $\mathrm{\delta_{Res}}$ is the correction factor whose details can be found in Ref.~\cite{RCorr-2018}.  
Panels (a) and (b) show the $R(\Delta S_2^{''})$ distributions from EBE-AVFD events of 30-40\% Ru+Ru and Zr+Zr collisions, respectively, at $\sqrt{s_{\rm NN}} = 200$ GeV with different $n_{5}/s$ inputs.
As $n_{5}/s$ increases, the $R(\Delta S_2^{''})$ distribution becomes more concave, qualitatively representing more CME contributions. To quantify the distribution shape, the Gaussian width ($\sigma_{R2}$) is obtained by fitting each $R(\Delta S_2^{''})$ distribution with an inverse Gaussian function, and the resultant  $\sigma^{-1}_{R2}$ values are listed in panel (c), increasing with $n_{5}/s$. 
The $\sigma^{-1}_{R2}$ ratios between Ru+Ru and Zr+Zr are shown in panel (d). According to Eq.~\ref{eq:relation2}, $\sigma^{-1}_{R2}$ is proportional to $\sqrt{\Delta\gamma_{112}} \propto [(\sigma^{-2}_{R2})^{\rm CME}+(\sigma^{-2}_{R2})^{\rm BG}]^{1/2}$, and hence the $\sigma^{-1}_{R2}$ ratio is expected to follow a quadratic trend vs $n_5/s$, assuming the CME signal is small compared with the background.
We fit the $\sigma^{-1}_{R2}$ ratios with a $2^{\rm nd}$-order polynomial function starting from (0, 1), though
the statistical uncertainties are large. The significance values of these ratios are stored in Table~\ref{tab:significance} for later discussions.

\begin{figure}
\hspace{-1.2cm}
\centering
\subfigure{\includegraphics[width=6.2cm]{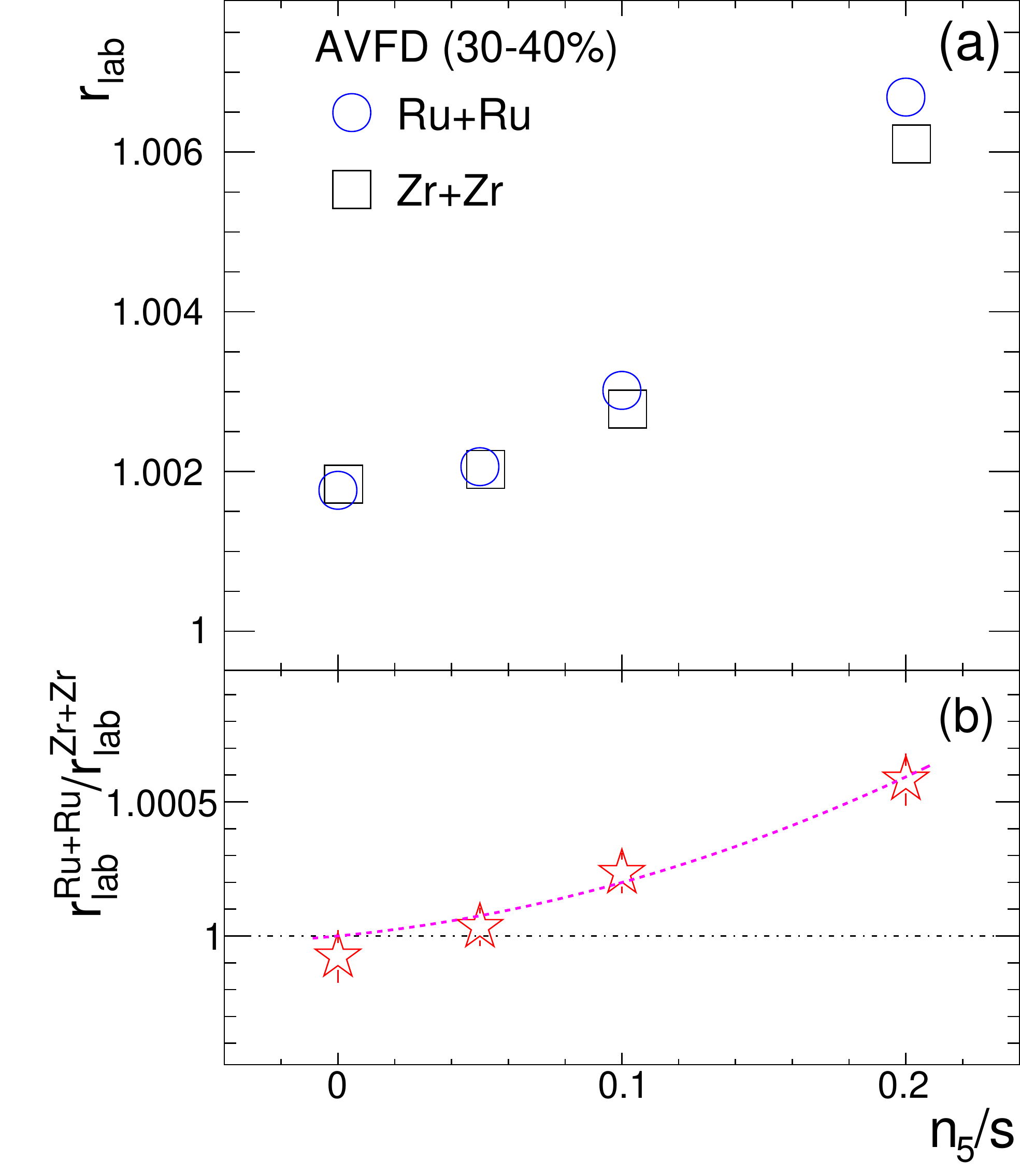}}
\subfigure{\includegraphics[width=6.2cm]{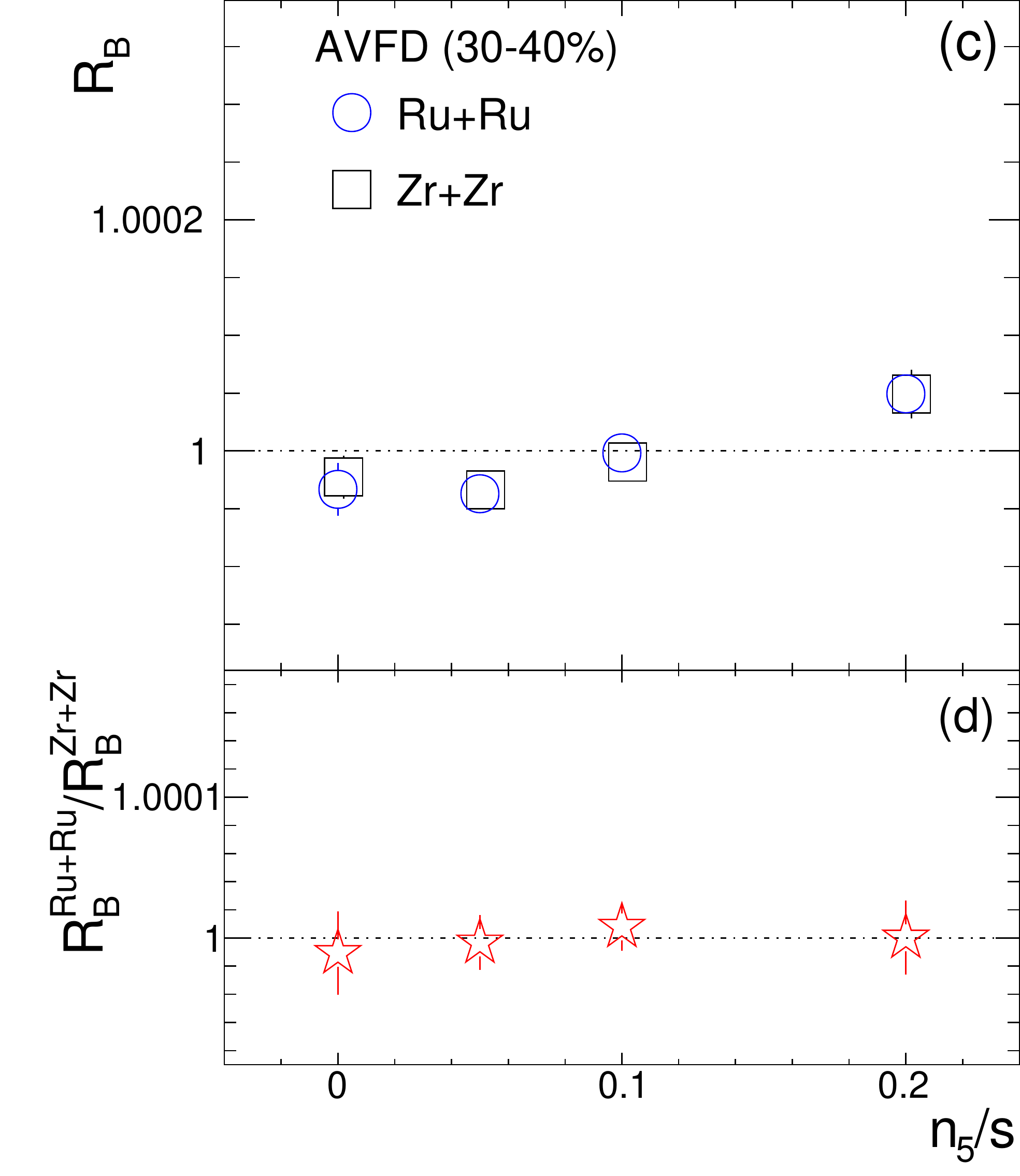}}
\caption{ $r_{\mathrm{lab}}$ (a) and $R_{\mathrm{B}}$ (c) as  function of $n_{5}/s$ from the EBE-AVFD model for 30-40\% Ru+Ru and Zr+Zr collisions at $\sqrt{s_{\rm NN}} =200$ GeV, with their ratios between Ru+Ru and Zr+Zr in panels (b) and (d), respectively.
In panel (b), the $2^{\rm nd}$-order-polynomial fit function demonstrates the rising trend starting from (0, 1).}
\hspace{1.43cm}\label{fig:SBF_isobar}
\end{figure}

Figure~\ref{fig:SBF_isobar}  presents the sensitivity study for the signed balance functions. This approach is not part of the STAR blind analysis, but follows the same procedure as used in the Quark Matter 2019 Conference proceedings~\cite{Yufu-2020}. 
The observables
$r_\mathrm{lab}$ and $R_{\rm B}$
as defined in Eqs.~\ref{rlab} and \ref{eq:RB} (with $p_T$ weighting), respectively, are exhibited in panels (a) and (c) as function of 
$n_{5}/s$ from the EBE-AVFD model for 30-40\% Ru+Ru and Zr+Zr collisions at $\sqrt{s_{\rm NN}} =200$ GeV. The corresponding 
ratios between Ru+Ru and Zr+Zr are shown in panels (b) and (d), respectively.
$r_\mathrm{lab}$ increases with the CME signal in each isobaric collision. According to Eqs.~\ref{rlab} and \ref{eq:relation3} and the  core-component
comparisons, $r_\mathrm{lab}$ is related to $\sqrt{\Delta\gamma_{112}}$, and therefore the $r_\mathrm{lab}$ ratio between the two systems should roughly obey a $2^{\rm nd}$-order polynomial function that starts from (0, 1). This relation is demonstrated with the corresponding fit in Fig.~\ref{fig:SBF_isobar}(b).
Panel (d) does not show a clear trend for 
the ratio of $R_{\rm B}^{\rm Ru+Ru}/R_{\rm B}^{\rm Zr+Zr}$, which is not a complete surprise:
$R_B$ looks for a higher-order effect in the difference between $r_\mathrm{lab}$ and $r_\mathrm{rest}$, and thus requires much more statistics than $r_\mathrm{lab}$. 

The sensitivity in terms of the statistical significance of $(O^{\rm Ru+Ru}/O^{\rm Zr+Zr} -1)$ is listed in Table ~\ref{tab:significance} for the observables, $\Delta\gamma_{112}$, $\Delta\delta$, $\kappa_{112}$, $r_{\rm lab}$ and $\sigma_{R2}^{-1}$.
This table serves as a reference point to interpret  the STAR data of the isobaric collisions. 
Note that opposite to other observables, the $\Delta\delta$ ratio is supposed to be  lower than unity in presence of the CME. Therefore, the more negative the statistical significance of $(\Delta\delta^{\rm Ru+Ru}/\Delta\delta^{\rm Zr+Zr}-1)$, the more sensitive this observable is to the CME signal. The high sensitivities of the $\Delta\delta$ ratio reported in Table~\ref{tab:significance} could be a special feature of EBE-AVFD, instead of a universal truth, which awaits verification/falsification of real data.
In general, 
$\kappa_{112}$ roughly doubles the sensitivity of $\Delta\gamma_{112}$, which, as explained before, should be mostly due to the contribution of $\Delta\delta$, and needs to be tested by experimental data.

$\Delta\gamma_{112}$ and $r_{\rm lab}$ show similar significance values because of the approximate equivalence between them.
Note that neither the toy model nor the EBE-AVFD model takes into account the separate CME domains that, instead of merging into a global charge separation for the whole event, still move independently from each other in the fireball. Thus $r_\mathrm{lab}$ is expected by these models to respond to the CME signal in a  similar way as $\Delta\gamma_{112}$ that only deals with the azimuthal angle.
Should the isobaric-collision data show a better sensitivity of $r_\mathrm{lab}$ than that of $\Delta\gamma_{112}$, it may reveal the CME domains undergoing an incomplete hydrodynamic evolution due to its short duration.

\begin{center}
\begin{table}[tbh]
\centering
\caption{The statistical significance of $(O^{\rm Ru+Ru}/O^{\rm Zr+Zr} -1)$ for different experimental observables. Contrary to other observables, the $\Delta\delta$ ratio expects negative  significance values due to the CME signal. $N_\mathrm{event}$ means the number of events used for each isobaric system in the simulation.}
\resizebox{0.38\textwidth}{!}{ 
    \begin{tabular}{c|cccccc}
    \toprule
    $n_{5}/s$ & $N_\mathrm{event}$ & $\Delta \gamma_{112} $ & $\Delta\delta$ & $\kappa_{112}$ & $r_{\mathrm{lab}}$  & $ \sigma_{R2}^{-1}$ \\
    \hline\noalign{\smallskip}
    0    &   $2\times 10^8 $   &  -1.50  &  -2.89 & -1.21  &  -0.77 & 1.33 \\
    0.05 &    $4\times 10^8 $   &   0.62 &  -6.16 &  1.37  &   0.47 & 0.29 \\
    0.10 &    $4\times 10^8 $   &  1.91 &  -16.81 &  3.43  &   3.11 & 0.62 \\
    0.20 &     $2\times 10^8 $  & 7.73 &   -42.96   & 14.07  &   5.96 & 1.84 \\
    \bottomrule
    \end{tabular}
    }
    \label{tab:significance}
\end{table}
\end{center}

In the analysis of the $R$ correlator using the STAR frozen code, $\sigma_{R2}^{-1}$ yields lower significance than other observables, and this is worth to note in anticipation of the results from the STAR blind analysis.  However,
this is largely due to two factors in this particular implementation.
First, this analysis uses the sub event plane instead of the full event plane as in the $\Delta\gamma_{112}$  analysis, which leads to worse event plane resolutions and hence larger statistical uncertainties.
Second, the particles of interest in the $R(\Delta S_2)$ analysis come from narrower kinematic regions than other analyses, which further enlarges its statistical errors and reduces its sensitivities.
When we repeat
the calculations of the $R$ correlator and the $\gamma_{112}$ correlator
both with the true reaction plane and with the same kinematic cuts  (not following the frozen code anymore), $\sigma_{R2}^{-1}$ and $\Delta\gamma_{112}$ do exhibit comparable significance values (not shown in this article), consistent with findings in Ref.~\cite{PhysRevC.103.034912}.
Therefore, this result, the toy model studies and Eqs.~\ref{eq:relation1} and \ref{eq:relation3} all give a coherent picture that on general grounds, 
the $\gamma_{112}$ correlator, the $R$ correlator and the signed balance functions have similar sensitivities, when used on the same set of particles.

\section{Summary}
Several experimental approaches have been developed to search for the CME in heavy-ion collisions.
In this paper, we focus on  three of them: the $\gamma_{112}$ correlator, the $R(\Delta S_2)$ correlator and the signed balance functions.
We have established the  relation between these methods via analytical derivation, and employed both simple Monte Carlo simulations and the EBE-AVFD model to verify the equivalence between the corecomponents of these observables.
Our study also supports the assumption that the CME signal and the background contributions can be linearly added up in such corecomponents.
For the observables of $\Delta\gamma_{112}$, $\Delta\delta$, $\kappa_{112}$, $r_{\rm lab}$ and $\sigma_{R2}^{-1}$, we have
extracted their sensitivities to the difference between Ru+Ru and Zr+Zr collisions at $\sqrt{s_{\rm NN}} = 200$ GeV from 30-40\%  central events generated by EBE-AVFD.
$\Delta\delta$ and $\kappa_{112}$ may render better sensitivities than other observables,  which could be a model-dependent feature instead of a universal truth, and needs to be further scrutinized by data.
The same significance level has been corroborated for $\Delta\gamma_{112}$, $r_{\rm lab}$ and $\sigma_{R2}^{-1}$, if put on an equal footing, but
the implementation details in the STAR frozen code can cause apparent differences in their sensitivities.
Therefore, this study provides a reference point to gauge the STAR isobaric-collision data.
 
\section*{Acknowledgement}
This effort of investigation was initiated in, and agreed upon by the CME focus group of the STAR Collaboration motivated by the desire to benchmark the various CME observables planned for use in the STAR isobar blind analysis against a well-established model EBE-AVFD. We thank the STAR Collaboration for the permitting the use of the corresponding isobar blind analysis frozen code in this simulation study, and the many collaborators who have contributed to previous STAR CME related studies besides the authors. We are especially grateful to the following people for their substantial help: Kenneth Barish, Helen Caines, Jinhui Chen, William Christie,  Frank Geurts, Huanzhong Huang, Hongwei Ke, William Llope, Xiaofeng Luo, Rongrong Ma, Yugang Ma, Bedanga Mohanty, Grigory Nigmatkulov, Lijuan Ruan, Ernst Sichtermann, Yuanfang Wu, Nu Xu, Zhangbu Xu, and Zhenyu Ye. 

We thank the STAR Collaboration, the RHIC Operations Group, and RCF at RHIC for their support.  This work is funded by the US Department of Energy under Grants No. DE-AC02-98CH10886, DE-FG02-89ER40531, 
DE-FG02-92ER40713, 
DE-FG02-88ER40424, 
DE-SC0012910,  
DE-SC0013391,  
DE-SC0020651, 
and by the National Natural Science Foundation of China under Contract No. 12025501, 
11905059, 12075085, 
the Strategic Priority Research Program of Chinese Academy of Science with Grant No. XDB34030200, 
the Fundamental Research Funds for the Central Universities under Grant No. CCNU19ZN019, the Ministry of Science and Technology (MoST) under Grant No. 2016YFE0104800 and the China Scholarship Council (CSC), Join Large-Scale Scientific Facility Funds of NSFC and CAS under Contracts No. U2032110, 
the U.S. Department of Energy, Office of Science, Office of Nuclear Physics, within the framework of the Beam Energy Scan Theory (BEST) Topical Collaboration,
the U.S. National Science Foundation under Grant No. PHY-1913729, 
the Natural Sciences and Engineering Research Council of Canada, 
 the Fonds de recherche du Qu\'ebec - Nature et technologies (FRQNT) through the Programmede Bourses d'Excellencepour \'Etudiants \'Etrangers (PBEEE). 

\appendix

\section{Derivation of $\sigma(\Delta B_y)$ and $\sigma(\Delta B_x)$}
\label{appendix1}

To estimate the RMS of $\Delta B_y$,  $\sigma(\Delta B_y)$, we first go through the following expansion,
\begin{eqnarray}
(N_{x(\alpha\beta)}-N_{x(\beta\alpha)})^2 &=& 
\Big\{\sum_{\alpha,\beta} {\rm Sign}[\sin(\phi^*_\alpha)-\sin(\phi^*_\beta)] \Big\}^2  \\
&=& \Big\{\sum_{\alpha,\beta} {\rm Sign}[\sin(\phi^*_\alpha)-\sin(\phi^*_\beta)] \Big\} \Big\{\sum_{\alpha',\beta'} {\rm Sign}[\sin(\phi^*_{\alpha'})-\sin(\phi^*_{\beta'})] \Big\} \nonumber \\
&=& \sum_{\alpha\neq\alpha'}\sum_{\beta\neq\beta'} {\rm Sign}[\sin(\phi^*_\alpha)-\sin(\phi^*_\beta)]\times{\rm Sign}[\sin(\phi^*_{\alpha'})-\sin(\phi^*_{\beta'})]  \nonumber \\
& &+ \sum_{\alpha}\sum_{\beta\neq\beta'} {\rm Sign}[\sin(\phi^*_\alpha)-\sin(\phi^*_\beta)]\times{\rm Sign}[\sin(\phi^*_{\alpha})-\sin(\phi^*_{\beta'})]  \nonumber \\
& &+ \sum_{\alpha\neq\alpha'}\sum_{\beta} {\rm Sign}[\sin(\phi^*_\alpha)-\sin(\phi^*_\beta)]\times{\rm Sign}[\sin(\phi^*_{\alpha'})-\sin(\phi^*_{\beta})]  
+ \sum_{\alpha}\sum_{\beta} 1. \label{eq:A2}
\end{eqnarray}
Then the average of each term in Eq.~\ref{eq:A2} is computed separately in a way similar to Eq.~\ref{eq:integral_result}.
The first one reads
\begin{eqnarray}
& &\Big\langle\sum_{\alpha\neq\alpha'}\sum_{\beta\neq\beta'} {\rm Sign}[\sin(\phi^*_\alpha)-\sin(\phi^*_\beta)]\times{\rm Sign}[\sin(\phi^*_{\alpha'})-\sin(\phi^*_{\beta'})]\Big\rangle  \nonumber \\
&=& N_\alpha(N_\alpha-1)N_\beta(N_\beta-1)\times
\Big\{
\int_{-\pi/2}^{\pi/2} 
\Big[\int_{-\pi-\phi^*_\alpha}^{\phi^*_\alpha} \frac{dN}{d\phi^*_\beta}d\phi^*_\beta-\int^{\pi-\phi^*_\alpha}_{\phi^*_\alpha} \frac{dN}{d\phi^*_\beta}d\phi^*_\beta\Big]\frac{dN}{d\phi^*_\alpha}d\phi^*_\alpha \nonumber \\
&  & \:\:\:\:\:\:\:\:\:\:\:\:\:\:\:\:\:\:\:\:\:\:\:\:\:\:\:\:\:\:\:\:\:\:\:\:\:\:\:\:\:\:\:\:\:\:\:\:\:\:\:\:\:\: +\int_{\pi/2}^{3\pi/2}
\Big[\int_{\pi-\phi^*_\alpha}^{\phi^*_\alpha} \frac{dN}{d\phi^*_\beta}d\phi^*_\beta-\int^{3\pi-\phi^*_\alpha}_{\phi^*_\alpha} \frac{dN}{d\phi^*_\beta}d\phi^*_\beta\Big]\frac{dN}{d\phi^*_\alpha}d\phi^*_\alpha
\Big\}^2 \nonumber \\
&=& N_\alpha(N_\alpha-1)N_\beta(N_\beta-1)\Big[\frac{8}{\pi^2}(1+\frac{2}{3}v_2)(a_{1,\alpha}-a_{1,\beta}) \Big]^2.
\label{eq:A3}
\end{eqnarray}
The second term becomes
\begin{eqnarray}
& &\Big\langle\sum_{\alpha}\sum_{\beta\neq\beta'} {\rm Sign}[\sin(\phi^*_\alpha)-\sin(\phi^*_\beta)]\times{\rm Sign}[\sin(\phi^*_{\alpha})-\sin(\phi^*_{\beta'})]\Big\rangle  \nonumber \\
&=& N_\alpha N_\beta(N_\beta-1)\times
\Big\{
\int_{-\pi/2}^{\pi/2} 
\Big[\int_{-\pi-\phi^*_\alpha}^{\phi^*_\alpha} \frac{dN}{d\phi^*_\beta}d\phi^*_\beta-\int^{\pi-\phi^*_\alpha}_{\phi^*_\alpha} \frac{dN}{d\phi^*_\beta}d\phi^*_\beta\Big]^2\frac{dN}{d\phi^*_\alpha}d\phi^*_\alpha \nonumber \\
&  & \:\:\:\:\:\:\:\:\:\:\:\:\:\:\:\:\:\:\:\:\:\:\:\:\:\:\:\:\:\:\:\:\:\:\:\: +\int_{\pi/2}^{3\pi/2}
\Big[\int_{\pi-\phi^*_\alpha}^{\phi^*_\alpha} \frac{dN}{d\phi^*_\beta}d\phi^*_\beta-\int^{3\pi-\phi^*_\alpha}_{\phi^*_\alpha} \frac{dN}{d\phi^*_\beta}d\phi^*_\beta\Big]^2\frac{dN}{d\phi^*_\alpha}d\phi^*_\alpha
\Big\} \nonumber \\
&=& N_\alpha N_\beta(N_\beta-1)\Big[\frac{1}{3}-\frac{8}{\pi^2}(1+v_2)a_{1,\beta}(a_{1,\alpha}-a_{1,\beta}) \Big].
\label{eq:A4}
\end{eqnarray}
Similarly, the third term gives
\begin{eqnarray}
& &\Big\langle\sum_{\alpha\neq\alpha'}\sum_{\beta} {\rm Sign}[\sin(\phi^*_\alpha)-\sin(\phi^*_\beta)]\times{\rm Sign}[\sin(\phi^*_{\alpha'})-\sin(\phi^*_{\beta})]\Big\rangle  \nonumber \\
&=& N_\alpha(N_\alpha-1) N_\beta\Big[\frac{1}{3}+\frac{8}{\pi^2}(1+v_2)a_{1,\alpha}(a_{1,\alpha}-a_{1,\beta}) \Big].
\label{eq:A5}
\end{eqnarray}
The last term is simply
\begin{equation}
\sum_{\alpha}\sum_{\beta} 1 = N_\alpha N_\beta. 
\label{eq:A6}
\end{equation}
According to the definition in Eq.~\ref{eq:by}, we have
\begin{equation}
\sigma^2(\Delta B_y) = \frac{M^2}{N_+^2 N_-^2} \langle  (N_{x(+-)}-N_{x(-+)})^2  \rangle.
\label{eq:A7}
\end{equation}
For simplicity, we assume $N_+ = N_- = M/2 \gg 1$ and $v_2 \ll 1$, and put Eqs.~\ref{eq:A3}, \ref{eq:A4}, \ref{eq:A5} and \ref{eq:A6} into Eq.~\ref{eq:A7}, so that
\begin{eqnarray}
\sigma^2(\Delta B_y) &\approx& \frac{4(M+1)}{3}+ \frac{64M^2}{\pi^4}\Big[(1+\frac{2}{3}v_2)^2+\frac{\pi^2(1+v_2)}{4M}\Big](a_{1,+}-a_{1,-})^2 \nonumber \\
&\approx& \frac{4M}{3}+ \frac{64M^2}{\pi^4}(1+ \frac{4}{3}v_2)(a_{1,+}-a_{1,-})^2.    
\end{eqnarray}
In a similar way, we also reach
\begin{eqnarray}
\sigma^2(\Delta B_x) &\approx& \frac{4(M+1)}{3}+ \frac{64M^2}{\pi^4}\Big[(1-\frac{2}{3}v_2)^2+\frac{\pi^2(1-v_2)}{4M}\Big](v_{1,+}-v_{1,-})^2 \nonumber \\
&\approx& \frac{4M}{3}+ \frac{64M^2}{\pi^4}(1- \frac{4}{3}v_2)(v_{1,+}-v_{1,-})^2.    
\end{eqnarray}
Then the difference between $\sigma^2(\Delta B_y)$ and $\sigma^2(\Delta B_x)$ will render
\begin{eqnarray}
\sigma^2(\Delta B_y) - \sigma^2(\Delta B_x) &\approx& \frac{128M^2}{\pi^4}\Big[(1+\frac{\pi^2}{4M})\Delta\gamma_{112}-(\frac{4}{3}+\frac{\pi^2}{4M})v_2\Delta\delta\Big] \\
&\approx& \frac{128M^2}{\pi^4}(\Delta\gamma_{112}-\frac{4}{3}v_2\Delta\delta).
\end{eqnarray}

{}

\end{document}